\documentclass[12pt,a4paper]{article}

\usepackage[utf8]{inputenc}
\usepackage[T1]{fontenc}
\usepackage{amsmath,amssymb,amsfonts}
\usepackage{bm}
\usepackage{graphicx}
\usepackage[colorlinks,citecolor=blue,urlcolor=blue,linkcolor=blue]{hyperref}
\usepackage{color}
\usepackage{enumerate}
\usepackage[margin=2.5cm]{geometry}

\graphicspath{{./figures/}}

\newcommand{\iu}{\mathrm{i}\mkern1mu}

\newcommand{\indf}{\mathbf{1}}

\newcommand{\Hyper}{\mathcal{H}}
\newcommand{\Vset}{\mathcal{V}}
\newcommand{\Eset}{\mathcal{E}}
\newcommand{\strategy}{a}
\newcommand{\info}{\mu}
\newcommand{\score}{U}
\newcommand{\payoff}{u}
\newcommand{\pol}{m}
\newcommand{\temp}{\Gamma}
\newcommand{\alphacrit}{\alpha_c}
\newcommand{\avg}[1]{\langle #1 \rangle}

\newcommand{\sgn}{\mathrm{sgn}}
\newcommand{\var}{\mathrm{Var}}

\newcommand{\dv}[2]{\frac{d #1}{d #2}}
\newcommand{\pdv}[2]{\frac{\partial #1}{\partial #2}}
\newcommand{\pdvtwo}[3]{\frac{\partial^2 #1}{\partial #2\,\partial #3}}

\begin{document}


\thispagestyle{empty}

\begin{center}

{\LARGE\bfseries Hypergraph Minority Game with Local Hyperedge Payoffs}\\[24pt]

{\large
Yihang Zhu\textsuperscript{1},
Fanyuan Meng\textsuperscript{1,*}\\[18pt]
}

{\footnotesize
\textsuperscript{1} Alibaba Research Center for Complexity Sciences, Hangzhou Normal University, Hangzhou 311121, Zhejiang, China

\textsuperscript{*} Corresponding author: fanyuan.meng@hotmail.com
}
\end{center}

\vspace{12pt}

\begin{abstract}
We provide a theoretical derivation of the Hypergraph Minority Game with Local Hyperedge Payoffs (HMG-L), in which $N$ adaptive agents compete simultaneously in multiple overlapping groups modeled as hyperedges of a static hypergraph $\Hyper=(\Vset,\Eset)$. Each hyperedge constitutes an independent local minority game, and agents accumulate payoffs across all groups to which they belong. We derive the continuum-time limit of the score dynamics, from which we obtain a set of coupled nonlinear stochastic differential equations for the agents' strategy polarization variables. The deterministic drift is shown to derive from a global cost function that generalizes the standard Minority Game Hamiltonian to hypergraph-structured interactions. We perform a sparse-annealed replica analysis of the stationary state for the case of a $k$-uniform, $d$-regular random hypergraph, obtaining the saddle-point equations within the replica-symmetric ansatz, an explicit replicon stability criterion, and Bethe/cavity equations for sparse corrections. The leading sparse-regime transition occurs on a critical surface $\alphacrit(k,d)$, while the globally coupled MG value $\alphacrit\simeq0.3374$ is recovered only in the separate single-hyperedge limit. We derive expressions for the order parameters---global volatility $\sigma^2$, predictability $\theta$, hyperedge frustration $F_e$, and frozen fraction $\phi$---and discuss their scaling behavior near criticality. The Fokker-Planck equation governing finite-$N$ fluctuations is presented, and the noise covariance matrix is computed from the hypergraph structure. Limiting cases ($k\to N$, $k\to2$, $d\to\infty$) are analyzed in detail, establishing connections to the standard MG, networked MG, and parallel MG models.
\end{abstract}

\noindent\textbf{Keywords:} Minority game; Hypergraph; Higher-order interactions; Replica method; Phase transition; Spin glass.

\vspace{12pt}


\section{Introduction}

The Minority Game (MG), introduced by Challet and Zhang~\cite{ChalletZhang97,ChalletZhang98} and inspired by Arthur's El Farol Bar problem~\cite{Arthur94}, has become a paradigmatic model in the statistical physics of complex adaptive systems. It describes a population of $N$ heterogeneous agents who repeatedly choose between two alternatives, with those in the minority group winning each round. Despite its deceptive simplicity---each agent holds a small set of randomly drawn look-up tables (strategies) mapping a public information signal to an action, and selects the strategy with the highest accumulated virtual score---the MG exhibits a remarkably rich collective behavior. The key control parameter is $\alpha=P/N$, the ratio of the size of the information space $P$ to the number of agents. As $\alpha$ varies, the system undergoes a phase transition at $\alpha_c\simeq0.3374$~\cite{ChalletMarsiliZecchina00,SavitManucaRiolo99}, separating a symmetric (crowded, herding) phase at small $\alpha$ from an asymmetric (frozen, predictable) phase at large $\alpha$.

The MG owes its enduring appeal to a fortunate confluence: it is simultaneously simple enough to admit an exact analytical solution via replica and generating functional techniques borrowed from spin-glass theory~\cite{ChalletMarsiliZecchina00, MarsiliChallet01, HeimelCoolen01, CoolenHeimelSherrington01, ChalletMarsiliZhang05}, while being rich enough to capture stylized facts of financial markets---fat tails, volatility clustering, and the emergence of market efficiency~\cite{ChalletChessaMarsiliZhang01}. These features have motivated an extensive literature of generalizations, including agents with market impact~\cite{ChalletMarsiliZecchina00}, evolutionary selection of strategies~\cite{HodNakar02,SysiAhoChakrabortiKaski04}, heterogeneous time scales, multi-choice extensions~\cite{ChowChau03}, contrarian agents~\cite{ZhongZhengZhengHui05}, rigorous game-theoretic analyses~\cite{RenaultScarlattiScarsini05}, laboratory experiments~\cite{LindeSonnemansTuinstra14}, reinforcement learning~\cite{ZhangDongLiuHuangHuangLai19}, and, crucially for the present work, the embedding of agents on complex networks~\cite{Anghel04}.

The networked MG, pioneered by Anghel et al.~\cite{Anghel04}, placed agents on an Erd\H{o}s--R\'enyi substrate graph where each agent follows the best-performing neighbor. This seemingly local interaction produced global consequences: a scale-free leadership hierarchy emerged spontaneously, and an optimal connectivity was found to minimize collective volatility. A theoretical explanation based on crowd-anticrowd cancellation was subsequently developed~\cite{HartJefferiesHuiJohnson01,LoChanHuiJohnson04}. Lee and Jeong~\cite{LeeJeong06} systematically extended this approach to regular lattices, scale-free networks, and Watts--Strogatz small-world graphs, demonstrating that network topology fundamentally alters the MG phase structure. More recently, the Parallel Minority Game~\cite{VemulaBiswas26} introduced multiple simultaneous minority competitions with overlapping agent participation, revealing that the conflict between local and global optimization creates a spin-glass-like landscape.

Yet all existing generalizations share a structural limitation: the interaction substrate is strictly dyadic. In Anghel's model, agents exchange information along pairwise edges. In the Parallel MG, agents compete in $D(D-1)/2$ independent pairwise matchings. Real-world resource allocation and social coordination problems, however, frequently involve \emph{genuinely higher-order interactions}, where three or more agents interact irreducibly within a single group. A committee vote, a multi-firm research consortium, a coalition of traders coordinating across correlated assets, or a species guild competing for overlapping resources---in each case, the group interaction cannot be decomposed into a sum of pairwise encounters without losing essential structure.

The natural mathematical framework for such systems is the hypergraph~\cite{Battiston20,Boccaletti23}, where hyperedges represent groups of arbitrary size. In recent years, hypergraphs have revolutionized our understanding of evolutionary game theory~\cite{AlvarezRodriguez20,BurgioMatamalasGomezArenas20}, synchronization, and contagion processes. Placing the public goods game on hypergraphs, for instance, revealed that cooperation thresholds shift dramatically as a function of hyperedge size distribution and hyperdegree heterogeneity~\cite{WangGao25,LinZhouFang25}. Iacopini et al.~\cite{IacopiniPetriBaronchelliBarrat19} showed that committed minorities overturn social conventions more effectively through higher-order interactions. However, to the best of our knowledge, the Minority Game has \emph{never} been formulated on a hypergraph substrate. This is a conspicuous lacuna: the MG framework captures resource competition and minority-induced coordination, which are arguably even more naturally suited to higher-order group structures than the cooperation dilemmas studied so far.

In this paper, we fill this gap by introducing and analytically deriving the main continuum and mean-field properties of the Hypergraph Minority Game with Local Hyperedge Payoffs (HMG-L). In our model, $N$ agents are placed on a $k$-uniform, $d$-regular random hypergraph $\Hyper=(\Vset,\Eset)$. Each hyperedge $e\in\Eset$ (of size $k$) constitutes an independent local minority game in which its $k$ members compete simultaneously. Agents accumulate payoffs across all $d$ hyperedges they belong to, and update their strategy scores by the price-taking virtual-score rule specified below. This construction introduces two new structural control parameters---the hyperedge size $k$ and the hyperdegree $d$---alongside the standard MG parameter $\alpha=P/N$.

We provide a multi-level analytical treatment of the HMG-L. First, we derive the continuum-time limit of the discrete score dynamics, obtaining a set of coupled nonlinear stochastic differential equations for the agents' strategy polarization variables $m_i(\tau)\in[-1,1]$. We show that the deterministic drift derives from a global cost function---an effective Hamiltonian---that generalizes the standard MG Hamiltonian to hypergraph-structured interactions. Second, we formulate the Fokker-Planck equation governing finite-$N$ fluctuations and compute the noise covariance matrix explicitly from the hypergraph incidence structure. Third, we perform a replica analysis of the stationary state within the replica-symmetric (RS) and sparse-annealed approximations, obtaining saddle-point equations that determine the phase structure in the locally tree-like regime. The corresponding sparse-regime critical surface is $\alpha_c(k,d)=2d\,\frac{k-1}{k^2}$,
which generalizes the single critical point of the standard MG to a two-dimensional manifold in the $(k,d)$ plane within this sparse approximation. The critical surface encodes two competing structural effects: increasing $k$ lowers the sparse-regime threshold, whereas increasing $d$ raises it by giving each agent more local games. The fully connected limit $k=N,d=1$ is not obtained by taking $k\to N$ in the sparse formula; it must be treated separately, where the standard MG value $\alpha_c\simeq0.3374$ is recovered from the single-global-hyperedge Hamiltonian.

We derive expressions for the order parameters---global volatility $\sigma^2$, normalized volatility $\nu$, predictability $\theta$, hyperedge frustration $F_e$, and frozen fraction $\phi$---in terms of the replica overlaps. The sparse-RS phase diagram suggests a testable hyperedge-size dependence of global volatility, while the associated critical exponents require numerical validation rather than being assumed a priori. We further compare uniform, heterogeneous, and scale-free hypergraph ensembles at the level of topology proxies, clarifying the role of topological heterogeneity. A coevolutionary extension (CMG-AH), in which the hypergraph adapts to game outcomes~\cite{GrossBlasius08}, is briefly outlined as a direction for future investigation.

The paper is organized as follows. Section~\ref{sec:model} defines the HMG-L model, including the hypergraph construction, agent strategies, payoff formulation, and score dynamics. Section~\ref{sec:continuum} derives the continuum-time limit and the effective Hamiltonian. Section~\ref{sec:meanfield} presents the mean-field theory and identifies the effective control parameter. Section~\ref{sec:langevin} formulates the Langevin description and computes the noise covariance. Section~\ref{sec:fp} gives the Fokker-Planck equation. Section~\ref{sec:replica} contains the replica analysis, leading to the saddle-point equations, critical surface, and RS stability criterion. Section~\ref{sec:order_param} defines the order parameters. Section~\ref{sec:phase_diagram} presents the phase diagram. Section~\ref{sec:limiting} analyzes limiting cases and the crossover to the standard MG. Section~\ref{sec:fss} discusses finite-size scaling and numerical validation. Section~\ref{sec:discussion} concludes with a summary and outlook.

\section{Model Definition}
\label{sec:model}

\subsection{Agent Population and Strategy Space}

We consider a population of $N$ agents, labeled $i=1,2,\ldots,N$. Each agent is endowed with $S$ strategies (typically $S=2$), which are fixed throughout the dynamics and constitute the quenched disorder of the model. For the strictly binary payoff rule below, ties in a hyperedge are resolved by the convention $\sgn(0)=0$; if one wants to exclude ties altogether, every hyperedge size $k_e$ should be odd. The linear-payoff formulation used in the analytical sections is well defined for both even and odd $k_e$.

The public information available to all agents at time $t$ is encoded in an integer variable $\info(t)\in\{1,2,\ldots,P\}$. Following the random-information variant of the Minority Game \cite{Cavagna99}, we take $\info(t)$ to be drawn uniformly and independently at each time step from the set $\{1,\ldots,P\}$. The parameter $P$ controls the complexity of the information space and is related to the memory length $M$ of the original MG formulation via $P=2^M$.

A strategy $s$ belonging to agent $i$ is a binary vector of length $P$:
\begin{equation}\label{eq:strategy}
\strategy_{s,i}=(\strategy_{s,i}^1,\strategy_{s,i}^2,\ldots,\strategy_{s,i}^P),\qquad \strategy_{s,i}^\info\in\{+1,-1\},
\end{equation}
where $\strategy_{s,i}^\info$ specifies the action (``buy'' $=+1$ or ``sell'' $=-1$) that strategy $s$ prescribes for information state $\info$. The strategies are drawn independently at the start of the game from an unbiased distribution:
\begin{equation}\label{eq:quenched}
\mathrm{Prob}(\strategy_{s,i}^\info=+1)=\mathrm{Prob}(\strategy_{s,i}^\info=-1)=\frac12,
\end{equation}
independently for each $i$, $s$, and $\info$. These strategy assignments are the quenched disorder over which we will later average.

\subsection{Hypergraph Construction}

\begin{figure}[htbp]
\centering
\includegraphics[width=0.8\textwidth]{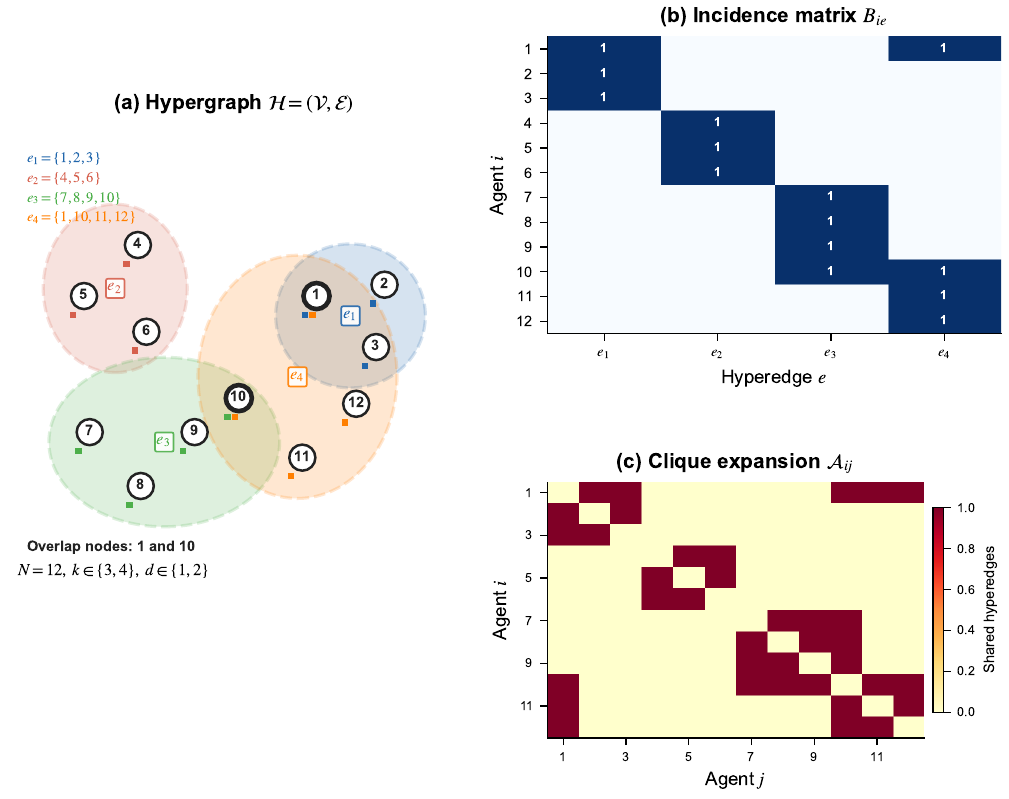}
\caption{\textbf{Hypergraph structure of the HMG-L model.}
(a) Schematic of a toy hypergraph with $N=12$ agents (circles) and $|\Eset|=4$ hyperedges (shaded regions) of sizes $k\in\{3,4\}$. The colored set labels list the exact membership of each hyperedge; colored squares next to a node mark the hyperedges containing that node, and thick node outlines identify overlap nodes. Each hyperedge $e$ constitutes an independent local minority game in which all member agents compete simultaneously.
(b) The incidence matrix $B_{ie}\in\{0,1\}$ encodes which agents belong to which hyperedges, with $\sum_i B_{ie}=k_e$ and $\sum_e B_{ie}=d_i$.
(c) The clique-expansion adjacency matrix $\mathcal{A}_{ij}=\sum_e B_{ie}B_{je}$ counts the number of hyperedges shared by agents $i$ and $j$, providing a measure of pairwise overlap induced by the higher-order structure.}
\label{fig:hypergraph_structure}
\end{figure}

The interaction structure among agents is encoded in a hypergraph $\Hyper=(\Vset,\Eset)$, where $\Vset=\{1,\ldots,N\}$ is the set of agents and $\Eset$ is a collection of hyperedges. The structure of the model is illustrated schematically in Fig.~\ref{fig:hypergraph_structure}.

We define the hyperdegree $d_i$ of agent $i$ as the number of hyperedges containing $i$:
\begin{equation}\label{eq:hyperdegree}
d_i=|\{e\in\Eset: i\in e\}|.
\end{equation}

For analytical tractability, we focus primarily on the ensemble of \emph{$k$-uniform, $d$-regular random hypergraphs}, in which:
\begin{itemize}
    \item All hyperedges have the same size: $k_e=k$ for all $e\in\Eset$;
    \item All agents have the same hyperdegree: $d_i=d$ for all $i\in\Vset$.
\end{itemize}
The total number of hyperedges is $|\Eset|=Nd/k$. Such hypergraphs can be sampled using the configuration model \cite{Chodrow20}: each agent $i$ is assigned $d$ ``stubs,'' and hyperedges are formed by randomly partitioning the $Nd$ stubs into groups of size $k$. We work in the sparse regime where $k,d\ll N$, and correlations between hyperedges are negligible (tree-like local structure) in the thermodynamic limit $N\to\infty$.

For extensions beyond the regular case, we allow the hyperedge size distribution $P(k)$ and the hyperdegree distribution $Q(d)$ to be arbitrary distributions with finite first and second moments.

\subsection{Hypergraph Incidence and Adjacency}

It is convenient to introduce the incidence matrix $B$ of the hypergraph, with entries:
\begin{equation}\label{eq:incidence}
B_{ie}=\begin{cases}
1, & \text{if agent } i\in e,\\
0, & \text{otherwise}.
\end{cases}
\end{equation}
The hyperdegree and hyperedge size are recovered as:
\begin{equation}
d_i=\sum_{e\in\Eset} B_{ie},\qquad k_e=\sum_{i\in\Vset} B_{ie}.
\end{equation}

The hypergraph adjacency matrix $\mathcal{A}$ (of the clique expansion) has entries:
\begin{equation}\label{eq:adjacency}
\mathcal{A}_{ij}=\sum_{e\in\Eset} B_{ie}B_{je},\qquad i\neq j,
\end{equation}
counting the number of hyperedges shared by agents $i$ and $j$. For a $k$-uniform, $d$-regular hypergraph, the average number of shared hyperedges between two randomly chosen agents is:
\begin{equation}
\avg{\mathcal{A}_{ij}}=\frac{d(k-1)}{N-1}\simeq\frac{d(k-1)}{N}.
\end{equation}

\subsection{Information Structure and Strategy Selection}

At each discrete time step $t=0,1,2,\ldots$:

\begin{enumerate}
    \item The public information $\info(t)\in\{1,\ldots,P\}$ is drawn uniformly at random.
    \item Each agent $i$ selects one of their $S$ strategies to play. The selection is governed by the Boltzmann (softmax) rule with inverse temperature $\temp>0$:
    \begin{equation}\label{eq:boltzmann}
    \mathrm{Prob}[s_i(t)=s]=\frac{\exp[\temp\,\score_{s,i}(t)]}{\sum_{s'=1}^S\exp[\temp\,\score_{s',i}(t)]},
    \end{equation}
    where $\score_{s,i}(t)$ is the virtual score (cumulative payoff) of strategy $s$ for agent $i$ at time $t$.
    \item For the selected strategy $s_i(t)$, agent $i$ plays the action $\strategy_{s_i(t),i}^{\info(t)}\in\{+1,-1\}$ in all hyperedges to which the agent belongs.
\end{enumerate}

\subsection{Payoff Formulation: Local Hyperedge Minority Games}

Within each hyperedge $e$, we define the \emph{local attendance imbalance}:
\begin{equation}\label{eq:local_attendance}
A_e(t)=\sum_{j\in e} a_{s_j(t),j}^{\info(t)}\;\in\{-k_e,-k_e+2,\ldots,k_e\}.
\end{equation}

The minority side within hyperedge $e$ is determined by the sign of $A_e(t)$: agents whose action matches $-\sgn[A_e(t)]$ are in the local minority and win. The payoff to agent $i$ from hyperedge $e$ is:
\begin{equation}\label{eq:local_payoff}
\payoff_i^e(t)=-a_{s_i(t),i}^{\info(t)}\cdot\sgn[A_e(t)].
\end{equation}

Agent $i$'s \emph{total payoff} at time $t$ is the sum of payoffs across all hyperedges they participate in:
\begin{equation}\label{eq:total_payoff}
\payoff_i(t)=\sum_{e\ni i}\payoff_i^e(t)=-\sum_{e\ni i}a_{s_i(t),i}^{\info(t)}\cdot\sgn[A_e(t)].
\end{equation}

Equivalently, we may work with the average per-hyperedge payoff $\payoff_i(t)/d_i$; the scaling does not affect the strategy selection dynamics provided all scores are scaled consistently.

\begin{figure}[t]
\centering
\includegraphics[width=\textwidth]{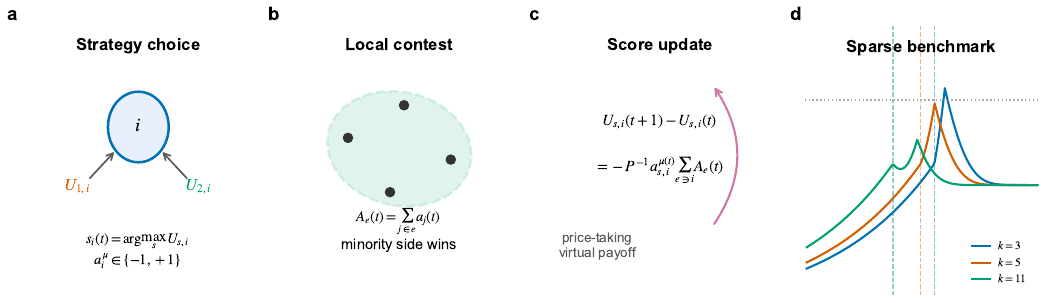}
\caption{\textbf{Model workflow of the HMG-L.}
(a) Each agent $i$ selects one of their $S$ strategies according to the Boltzmann rule based on virtual scores $U_{s,i}$.
(b) Within each hyperedge $e$, the local attendance $A_e=\sum_{j\in e}a_j$ determines the minority side $-{\rm sgn}(A_e)$; agents on the minority side win.
(c) Strategy scores are updated by the price-taking virtual payoff across all hyperedges the agent belongs to, creating a feedback loop between individual adaptation and collective outcomes.
(d) Compact sparse-closure benchmark for the normalized global volatility $\nu=\sigma^2/(4N)$, with vertical dashed lines marking $\alpha_c(k,d)$ for representative $k$ at fixed $d=3$.}
\label{fig:workflow}
\end{figure}

\subsection{Strategy Score Dynamics}

Each agent maintains a virtual score $\score_{s,i}(t)$ for each of their strategies. Following the standard MG prescription, the virtual score is updated by a \emph{price-taking} or \emph{batch} rule: each candidate strategy is evaluated against the realized aggregate attendance in the hyperedges containing the agent. This convention is the one that yields the standard MG Hamiltonian in the globally coupled limit:
\begin{equation}\label{eq:score_update}
\score_{s,i}(t+1)=\score_{s,i}(t)-\frac{1}{P}\,\strategy_{s,i}^{\info(t)}\sum_{e\ni i}\sgn[A_e(t)].
\end{equation}
If instead one uses the fully counterfactual attendance in hyperedge $e$,
\begin{equation}\label{eq:counterfactual}
A_e^{(s)}(t)=A_e(t)-a_{s_i(t),i}^{\info(t)}+a_{s,i}^{\info(t)}.
\end{equation}
then, for the linear payoff, the exact score-difference update contains $A_e(t)-a_{s_i(t),i}^{\info(t)}$ rather than $A_e(t)$. The difference is the agent's self-impact term; it is $O(1/k_e)$ relative to the local attendance for large hyperedges, but it changes the exact Hamiltonian. In the remainder of this derivation we therefore keep the price-taking update \eqref{eq:score_update}.

The factor $1/P$ in \eqref{eq:score_update} ensures that score increments are of order $1/P$, which is necessary for a well-defined continuum time limit. It is also conventional to consider the score \emph{difference} between strategies when $S=2$. Defining:
\begin{equation}\label{eq:score_diff}
y_i(t)=\frac{\score_{1,i}(t)-\score_{2,i}(t)}{2},
\end{equation}
the update rule becomes:
\begin{equation}\label{eq:score_diff_update}
y_i(t+1)=y_i(t)-\frac{1}{P}\,\xi_i^{\info(t)}\sum_{e\ni i}\sgn[A_e(t)],
\end{equation}
where we have introduced the \emph{strategy difference} variable:
\begin{equation}\label{eq:xi}
\xi_i^\info=\frac{a_{1,i}^\info-a_{2,i}^\info}{2}\in\{-1,0,+1\},
\end{equation}
with $\mathrm{Prob}(\xi_i^\info=\pm1)=1/4$, $\mathrm{Prob}(\xi_i^\info=0)=1/2$, independently for each $i$ and $\info$.

For the linear payoff case ($\sgn[A_e]\to A_e$, corresponding to proportional payoffs rather than strictly binary minority/majority), which we adopt henceforth for analytical tractability, the update simplifies to:
\begin{equation}\label{eq:score_diff_update_linear}
y_i(t+1)=y_i(t)-\frac{1}{P}\,\xi_i^{\info(t)}\sum_{e\ni i}A_e(t).
\end{equation}

\section{Continuum Time Limit}
\label{sec:continuum}

\subsection{Scaling Arguments}

Two key observations motivate the continuum time limit \cite{MarsiliChallet01}:

\begin{enumerate}
    \item \textbf{Small score increments:} From \eqref{eq:score_diff_update_linear}, a single microscopic update has magnitude $\Delta y_i=O(d\sqrt{k}/P)$ when different local attendances add incoherently and $A_e=O(\sqrt{k})$ by the Central Limit Theorem. For fixed $k,d$ and fixed $\alpha=P/N$, this is $O(1/N)$. If $k$ or $d$ grows with $N$, the continuum limit requires $d\sqrt{k}=o(P)$.
    \item \textbf{Long characteristic times:} Agents need to sample all $P$ information states to reliably assess their strategies. The natural time unit is therefore $P$ microscopic steps.
\end{enumerate}

We introduce the rescaled time:
\begin{equation}\label{eq:rescaled_time}
\tau=\frac{t}{P}.
\end{equation}
In the limit $N\to\infty$, $P\to\infty$ with $\alpha=P/N$ fixed, the discrete-time dynamics converge to a set of coupled stochastic differential equations.

\subsection{Strategy Polarization Variables}

For $S=2$ strategies per agent, the state of agent $i$ is fully characterized by the \emph{strategy polarization}:
\begin{equation}\label{eq:polarization}
\pol_i(\tau)=p_{1,i}(\tau)-p_{2,i}(\tau),
\end{equation}
where $p_{s,i}(\tau)=\mathrm{Prob}[s_i(\tau)=s]$ is the probability that agent $i$ plays strategy $s$ at rescaled time $\tau$. From the Boltzmann rule \eqref{eq:boltzmann}:
\begin{equation}\label{eq:polarization_boltzmann}
\pol_i(\tau)=\frac{e^{\temp \score_{1,i}}-e^{\temp \score_{2,i}}}{e^{\temp \score_{1,i}}+e^{\temp \score_{2,i}}}
               =\tanh[\temp\,y_i(\tau)].
\end{equation}
Thus $\pol_i\in[-1,1]$: $\pol_i=+1$ means agent $i$ deterministically plays strategy 1; $\pol_i=-1$ means they deterministically play strategy 2; $\pol_i=0$ means they randomize uniformly between the two.

The average action of agent $i$ in information state $\info$ is:
\begin{equation}\label{eq:avg_action}
\avg{a_i^\info}=\frac{1+\pol_i}{2}\,a_{1,i}^\info+\frac{1-\pol_i}{2}\,a_{2,i}^\info
               =\frac{a_{1,i}^\info+a_{2,i}^\info}{2}+\pol_i\,\xi_i^\info.
\end{equation}

For notational simplicity, we define the \emph{unbiased component} $\bar{a}_i^\info=(a_{1,i}^\info+a_{2,i}^\info)/2$, which takes values in $\{-1,0,+1\}$ and satisfies $\avg{\bar{a}_i^\info}=0$, $\avg{(\bar{a}_i^\info)^2}=1/2$. The average attendance in hyperedge $e$ for information state $\info$ is then:
\begin{equation}\label{eq:avg_attendance_hyperedge}
\avg{A_e^\info}=\sum_{j\in e}\avg{a_j^\info}
               =\sum_{j\in e}\bar{a}_j^\info+\sum_{j\in e}\pol_j\,\xi_j^\info.
\end{equation}

\subsection{Deterministic Drift}

The evolution of the score difference $y_i$ follows from \eqref{eq:score_diff_update_linear}. In the continuum limit, averaging over the random drawing of $\info$ and over the local fluctuations within hyperedges, we obtain the deterministic drift:
\begin{equation}\label{eq:drift_raw}
\dv{y_i}{\tau}=-\frac{1}{P}\sum_{\info=1}^P\xi_i^\info\sum_{e\ni i}\avg{A_e^\info}.
\end{equation}

Using $\pol_i=\tanh(\temp y_i)$, we derive the equation of motion for the polarization:
\begin{equation}\label{eq:dm_dtau}
\dv{\pol_i}{\tau}=\temp(1-\pol_i^2)\dv{y_i}{\tau}
                  =-\frac{\temp}{P}(1-\pol_i^2)\sum_{\info=1}^P\xi_i^\info\sum_{e\ni i}\avg{A_e^\info}.
\end{equation}

Inserting \eqref{eq:avg_attendance_hyperedge}:
\begin{equation}\label{eq:dm_dtau_expanded}
\dv{\pol_i}{\tau}=-\frac{\temp}{P}(1-\pol_i^2)\sum_{\info=1}^P\xi_i^\info
                 \left[\sum_{e\ni i}\sum_{j\in e}\bar{a}_j^\info
                 +\sum_{e\ni i}\sum_{j\in e}\pol_j\,\xi_j^\info\right].
\end{equation}

This is the central deterministic equation of motion for the HMG-L model. It describes how each agent's strategy polarization evolves in response to the aggregate behavior within the hyperedges they participate in.

\subsection{Effective Hamiltonian}

The dynamics \eqref{eq:dm_dtau_expanded} can be expressed as a gradient descent on a global cost function. Define the \emph{hyperedge Hamiltonian}:
\begin{equation}\label{eq:hamiltonian_local}
H_e[\bm{\pol}]=\frac{1}{P}\sum_{\info=1}^P\bigl(\avg{A_e^\info}\bigr)^2,
\end{equation}
which measures the squared attendance imbalance in hyperedge $e$, averaged over information states. The total Hamiltonian is:
\begin{equation}\label{eq:hamiltonian_total}
H[\bm{\pol}]=\sum_{e\in\Eset}H_e[\bm{\pol}]
             =\frac{1}{P}\sum_{e\in\Eset}\sum_{\info=1}^P
               \left(\sum_{j\in e}\bar{a}_j^\info+\sum_{j\in e}\pol_j\xi_j^\info\right)^2.
\end{equation}

One can verify that:
\begin{equation}\label{eq:gradient_relation}
\dv{\pol_i}{\tau}=-\frac{\temp}{2}(1-\pol_i^2)\pdv{H}{\pol_i},
\end{equation}
so that the dynamics constitute a constrained gradient descent on $H[\bm{\pol}]$ (the factor $(1-\pol_i^2)$ ensures $\pol_i$ remains in $[-1,1]$). The stationary states of the deterministic dynamics correspond to local minima of $H$, with the constraint $|\pol_i|\le1$.

For the globally coupled MG ($k=N$, single hyperedge), $H[\bm{\pol}]$ reduces to the standard MG Hamiltonian \cite{ChalletMarsiliZecchina00}:
\begin{equation}
H_{\mathrm{MG}}[\bm{\pol}]=\frac{1}{P}\sum_{\info=1}^P\Bigl(\sum_{j=1}^N(\bar{a}_j^\info+\pol_j\xi_j^\info)\Bigr)^2.
\end{equation}

\subsection{Expansion in Hypergraph Structure}

To make the hypergraph dependence explicit, we expand \eqref{eq:dm_dtau_expanded}. For a $k$-uniform, $d$-regular hypergraph, each agent belongs to exactly $d$ hyperedges, and each hyperedge contains exactly $k$ agents. Summing over hyperedges containing $i$ yields:
\begin{equation}\label{eq:expanded_drift}
\begin{aligned}
\dv{\pol_i}{\tau}=-\frac{\temp}{P}(1-\pol_i^2)\sum_{\info=1}^P\xi_i^\info
\Bigg[&d\,\bar{a}_i^\info+d\,\pol_i\xi_i^\info \\
&+\sum_{e\ni i}\sum_{j\in e\setminus\{i\}}(\bar{a}_j^\info+\pol_j\xi_j^\info)\Bigg].
\end{aligned}
\end{equation}

The first two terms represent the self-interaction of agent $i$ within their own hyperedges. The third term couples agent $i$ to their hypergraph neighbors. Defining the \emph{hyperedge-averaged neighbor field}:
\begin{equation}\label{eq:neighbor_field}
h_i^{\mathrm{neigh}}(\info)=\frac{1}{d}\sum_{e\ni i}\frac{1}{k-1}\sum_{j\in e\setminus\{i\}}(\bar{a}_j^\info+\pol_j\xi_j^\info),
\end{equation}
the drift becomes:
\begin{equation}\label{eq:drift_compact}
\dv{\pol_i}{\tau}=-\frac{\temp d}{P}(1-\pol_i^2)\sum_{\info=1}^P\xi_i^\info
                 \Bigl[\bar{a}_i^\info+\pol_i\xi_i^\info+(k-1)h_i^{\mathrm{neigh}}(\info)\Bigr].
\end{equation}

\subsection{Global and Local Decomposition}

It is useful to separate the dynamics into global (mean-field) and local (hyperedge-fluctuation) components. Define the \emph{global attendance} for information state $\info$:
\begin{equation}\label{eq:global_attendance}
A^\info=\sum_{j=1}^N(\bar{a}_j^\info+\pol_j\xi_j^\info),
\end{equation}
which is the total attendance imbalance across all agents (as in the standard MG). The hyperedge-averaged neighbor field can be decomposed as:
\begin{equation}\label{eq:neighbor_decomp}
h_i^{\mathrm{neigh}}(\info)=\underbrace{\frac{1}{N}\sum_{j=1}^N(\bar{a}_j^\info+\pol_j\xi_j^\info)}_{\text{global mean } A^\info/N}
                            +\underbrace{\delta h_i^{\mathrm{neigh}}(\info)}_{\text{local fluctuation}}.
\end{equation}

The local fluctuation $\delta h_i^{\mathrm{neigh}}$ arises from the fact that agent $i$'s hypergraph neighborhood is a finite sample of size $d(k-1)$ from the population. By the Central Limit Theorem, for large $d(k-1)$:
\begin{equation}
\delta h_i^{\mathrm{neigh}}(\info)\sim\mathcal{N}\!\left(0,\frac{\sigma_h^2}{d(k-1)}\right),
\end{equation}
where $\sigma_h^2=\var_j(\bar{a}_j^\info+\pol_j\xi_j^\info)$.

Substituting the decomposition \eqref{eq:neighbor_decomp} into \eqref{eq:drift_compact}:
\begin{equation}\label{eq:drift_decomposed}
\begin{aligned}
\dv{\pol_i}{\tau}=&-\frac{\temp d}{P}(1-\pol_i^2)\sum_{\info=1}^P\xi_i^\info
                   \Bigl[\bar{a}_i^\info+\pol_i\xi_i^\info+(k-1)\frac{A^\info}{N}\Bigr] \\
                 &-\frac{\temp d(k-1)}{P}(1-\pol_i^2)\sum_{\info=1}^P\xi_i^\info\,\delta h_i^{\mathrm{neigh}}(\info).
\end{aligned}
\end{equation}

The first line contains the \emph{mean-field dynamics} (coupling agent $i$ to the global attendance $A^\info$). The second line contains the \emph{hyperedge fluctuation term}, which vanishes as $d(k-1)\to\infty$ (the fully connected limit).

\section{Mean-Field Theory}
\label{sec:meanfield}

\subsection{Mean-Field Closure}

In the thermodynamic limit $N\to\infty$ with fixed $k$ and $d$, the hypergraph has a locally tree-like structure, and correlations between distinct hyperedges are negligible. The mean-field closure consists of approximating the neighbor field $h_i^{\mathrm{neigh}}(\info)$ by its average over the hypergraph ensemble, conditional on the global state of the system.

For a $k$-uniform, $d$-regular hypergraph, we make the ansatz:
\begin{equation}\label{eq:mf_ansatz}
\sum_{e\ni i}\sum_{j\in e\setminus\{i\}}(\bar{a}_j^\info+\pol_j\xi_j^\info)
\approx d(k-1)\,\avg{\bar{a}^\info+\pol\xi^\info}_{\mathrm{pop}},
\end{equation}
where $\avg{\cdot}_{\mathrm{pop}}$ denotes the population average over all agents. The population average of the unbiased component vanishes only after the quenched disorder average; for a fixed sample it is an $O(N^{-1/2})$ random field. The exact finite-$N$ mean-field identity is therefore:
\begin{equation}\label{eq:mf_ansatz2}
\avg{\bar{a}^\info+\pol\xi^\info}_{\mathrm{pop}}=\frac{1}{N}\sum_{j=1}^N(\bar{a}_j^\info+\pol_j\xi_j^\info)
                                               \equiv\langle\bar{a}^\info+\pol\xi^\info\rangle
                                               =\frac{A^\info}{N}.
\end{equation}

\subsection{Mean-Field Equation of Motion}

Inserting the mean-field ansatz into \eqref{eq:drift_compact} yields the closed equation:
\begin{equation}\label{eq:mf_dynamics}
\dv{\pol_i}{\tau}=-\frac{\temp d}{P}(1-\pol_i^2)\sum_{\info=1}^P\xi_i^\info
                 \Bigl[\bar{a}_i^\info+\pol_i\xi_i^\info+(k-1)\langle\bar{a}^\info+\pol\xi^\info\rangle\Bigr].
\end{equation}

This shows that, at the mean-field level, each agent couples to the population through the global attendance density $A^\info/N$, with an effective coupling strength proportional to $d(k-1)$.

For comparison with the standard MG, it is instructive to rewrite \eqref{eq:mf_dynamics} in terms of the global attendance $A^\info$ (see Fig.~\ref{fig:workflow} for a schematic summary of the complete model dynamics). Using the exact identity $A^\info=N\langle\bar{a}^\info+\pol\xi^\info\rangle$, we have:
\begin{equation}\label{eq:mf_dynamics_A}
\dv{\pol_i}{\tau}=-\frac{\temp d}{P}(1-\pol_i^2)\sum_{\info=1}^P\xi_i^\info
                 \Bigl[\bar{a}_i^\info+\pol_i\xi_i^\info+\frac{k-1}{N}A^\info\Bigr].
\end{equation}

\subsection{Effective Control Parameter}

Comparing \eqref{eq:mf_dynamics_A} with the standard MG equation of motion \cite{MarsiliChallet01}:
\begin{equation}
\dv{\pol_i}{\tau}\Big|_{\mathrm{MG}}=-\frac{\temp}{P}(1-\pol_i^2)\sum_{\info=1}^P\xi_i^\info\,A^\info,
\end{equation}
we see that the HMG-L introduces two modifications:

\begin{enumerate}
    \item A \textbf{prefactor} $d$, reflecting that each agent participates in $d$ simultaneous minority games;
    \item A \textbf{self-interaction term} $(\bar{a}_i^\info+\pol_i\xi_i^\info)$, which is of relative order $1/k$ compared to the neighbor coupling.
\end{enumerate}

The self-interaction term is smaller than the neighbor term by a factor $1/(k-1)$ when $k$ is large, but it should not be discarded for small hyperedges. At the level of a leading large-$k$ mean-field estimate, the dominant effect of the hypergraph is to rescale the effective coupling. Keeping only the hyperdegree prefactor gives the rough control parameter:
\begin{equation}\label{eq:alpha_eff}
\alpha_{\mathrm{eff}}=\frac{\alpha}{d},
\end{equation}
where $\alpha=P/N$ is the standard MG control parameter.

For large $k$, the $k-1$ factor in \eqref{eq:mf_dynamics_A} further enhances the neighbor coupling. The corresponding large-$k$ estimate is:
\begin{equation}\label{eq:alpha_eff_full}
\alpha_{\mathrm{eff}}=\frac{\alpha}{d(k-1)}.
\end{equation}

This scaling is only a heuristic comparison with the standard MG. It suggests the estimate:
\begin{equation}\label{eq:alpha_crit_surface}
\alphacrit^{\mathrm{MF}}(k,d)=d(k-1)\,\alphacrit^{\mathrm{MG}},
\end{equation}
where $\alphacrit^{\mathrm{MG}}\simeq0.3374$ is the critical value for the globally coupled MG. The sparse-annealed RS calculation in Sec.~\ref{sec:replica} gives a different, pair-count-normalized threshold \eqref{eq:alpha_c}; the two formulas correspond to different normalizations of the local attendance and should not be conflated.

\subsection{Stationary States and Stability}

The stationary states of the mean-field dynamics \eqref{eq:mf_dynamics} satisfy $\dv{\pol_i}{\tau}=0$ for all $i$, which requires either $\pol_i=\pm1$ (frozen agents) or the vanishing of the drift term.

For $\alpha>\alphacrit(k,d)$, the system is in the \emph{asymmetric phase} where a finite fraction $\phi$ of agents are frozen ($\pol_i=\pm1$) and the remaining agents have continuous $\pol_i$ values. The order parameter $Q=(1/N)\sum_i\pol_i^2<1$.

For $\alpha<\alphacrit(k,d)$, the system is in the \emph{symmetric phase} where no agents are frozen ($\phi=0$), the distribution of $\pol_i$ is continuous and symmetric about zero, and the attendance $A^\info$ vanishes on average for all $\info$. The response susceptibility may diverge as $\alpha\to\alphacrit$, depending on the stability of the RS solution.

For $\alpha\gg\alphacrit(k,d)$, the quenched strategies become effectively uncorrelated with the information patterns. Agents may freeze with random signs, but the aggregate attendance approaches the coin-tossing baseline $\sigma^2/N=1$ for actions $a_i=\pm1$. In the figures below we plot the rescaled volatility
\begin{equation}\label{eq:nu_definition}
\nu=\frac{\sigma^2}{4N},
\end{equation}
for which the independent-action baseline is $\nu=1/4$.

\section{Langevin Description and Fluctuations}
\label{sec:langevin}

\subsection{Noise Sources}

The deterministic drift derived in Sec.~\ref{sec:continuum} captures the average behavior. Fluctuations arise from three sources:

\begin{enumerate}
    \item \textbf{Sampling noise:} The information state $\info(t)$ is drawn randomly at each step. This produces temporal fluctuations in the score updates of relative order $1/\sqrt{P}$.
    \item \textbf{Hypergraph disorder:} The specific realization of the hypergraph introduces quenched fluctuations in the local neighborhood of each agent. These are of relative order $1/\sqrt{d(k-1)}$.
    \item \textbf{Finite-$N$ effects:} The discreteness of the agent population generates fluctuations in the global attendance of order $1/\sqrt{N}$.
\end{enumerate}

All three noise sources become Gaussian white noise in the continuum limit (by virtue of the Central Limit Theorem applied to the sum of many independent increments).

\subsection{Langevin Equation}

The full stochastic dynamics of the polarization variables are described by the Langevin equation:
\begin{equation}\label{eq:langevin}
\dv{\pol_i}{\tau}=\temp(1-\pol_i^2)\Bigl[F_i^{\mathrm{det}}(\bm{\pol})+\eta_i(\tau)\Bigr],
\end{equation}
where $F_i^{\mathrm{det}}$ is the deterministic force from \eqref{eq:mf_dynamics}:
\begin{equation}\label{eq:F_det}
F_i^{\mathrm{det}}(\bm{\pol})=-\frac{d}{P}\sum_{\info=1}^P\xi_i^\info
                             \Bigl[\bar{a}_i^\info+\pol_i\xi_i^\info+(k-1)\langle\bar{a}^\info+\pol\xi^\info\rangle\Bigr],
\end{equation}
For independent strategy choices at fixed $\bm{\pol}$,
\begin{equation}
\avg{\bigl(A_e^\info-\avg{A_e^\info}\bigr)\bigl(A_{e'}^\info-\avg{A_{e'}^\info}\bigr)}
=\sum_{\ell\in e\cap e'}(\xi_\ell^\info)^2(1-\pol_\ell^2),
\end{equation}
because only agents belonging to both hyperedges contribute to both attendance fluctuations. This expression is dimensionless, as are $A_e$, $\xi_i$, and $\pol_i$; hence $C_{ij}$ and the noise strength in \eqref{eq:noise_covariance} have the correct dimension of inverse rescaled time.
and $\eta_i(\tau)$ is a Gaussian noise term with zero mean and covariance:
\begin{equation}\label{eq:noise_covariance}
\avg{\eta_i(\tau)\,\eta_j(\tau')}=\frac{1}{P}\,C_{ij}(\bm{\pol})\,\delta(\tau-\tau').
\end{equation}

\subsection{Noise Covariance Matrix}

The noise covariance matrix $C_{ij}(\bm{\pol})$ encodes the correlations induced by the hypergraph structure. It is computed from the variance of the score increment:
\begin{equation}\label{eq:covariance_def}
C_{ij}(\bm{\pol})=\frac{1}{P}\sum_{\info=1}^P
                 \xi_i^\info\xi_j^\info\,
                 \sum_{e\ni i}\sum_{e'\ni j}
                 \avg{\bigl(A_e^\info-\avg{A_e^\info}\bigr)\bigl(A_{e'}^\info-\avg{A_{e'}^\info}\bigr)}.
\end{equation}

For a $k$-uniform, $d$-regular hypergraph, the covariance decomposes into diagonal and off-diagonal contributions:

\textbf{Diagonal ($i=j$):} Agent $i$ belongs to $d$ hyperedges. The fluctuations within each hyperedge are correlated because agent $i$'s own action contributes to all of them:
\begin{equation}\label{eq:C_diagonal}
C_{ii}(\bm{\pol})=\frac{1}{P}\sum_{\info=1}^P(\xi_i^\info)^2
                 \Biggl[\sum_{e\ni i}\sum_{\ell\in e}(\xi_\ell^\info)^2(1-\pol_\ell^2)
                 +d(d-1)(\xi_i^\info)^2(1-\pol_i^2)\Biggr].
\end{equation}

For a locally tree-like regular hypergraph and large $P$, using $\avg{(\xi_i^\info)^2}=1/2$, we obtain the annealed estimate:
\begin{equation}\label{eq:C_diagonal_simple}
C_{ii}(\bm{\pol})\simeq
\frac{d}{2}\left[\frac{1}{2}(1-\pol_i^2)+\frac{k-1}{2}(1-Q_{\partial i})\right]
\frac{d(d-1)}{4}(1-\pol_i^2),
\end{equation}
where $Q_{\partial i}$ is the mean of $\pol_\ell^2$ over neighbors of $i$ in the incident hyperedges. In a homogeneous mean-field closure $Q_{\partial i}$ is replaced by $Q=N^{-1}\sum_i\pol_i^2$.

\textbf{Off-diagonal ($i\neq j$):} Two agents $i$ and $j$ have correlated noise only if they share at least one hyperedge. Let $n_{ij}=|\{e:i,j\in e\}|$ be the number of hyperedges containing both agents. Then:
\begin{equation}\label{eq:C_offdiagonal}
C_{ij}(\bm{\pol})=\frac{1}{P}\sum_{\info=1}^P\xi_i^\info\xi_j^\info\,
                 \Biggl[\sum_{e\ni i}\sum_{e'\ni j}\sum_{\ell\in e\cap e'}(\xi_\ell^\info)^2(1-\pol_\ell^2)\Biggr].
\end{equation}

For a sparse regular hypergraph, $n_{ij}$ is Poisson-distributed with mean $d(k-1)/(N-1)\ll1$, so off-diagonal noise correlations are suppressed by $1/N$.

\subsection{Effective Temperature}

Comparing the Langevin equation \eqref{eq:langevin} with the gradient relation \eqref{eq:gradient_relation}, one obtains an equilibrium Gibbs form only after the diagonal, homogeneous covariance approximation $C_{ij}\simeq C\,\delta_{ij}$ and a matching fluctuation-dissipation normalization. With the conventional MG normalization this identifies an effective temperature:
\begin{equation}\label{eq:effective_temp}
T_{\mathrm{eff}}=\frac{1}{2\temp},
\end{equation}
in agreement with the standard MG result \cite{MarsiliChallet01}. Under this diagonal FDT approximation, the stationary distribution of the Langevin dynamics is of the Gibbs form:
\begin{equation}\label{eq:gibbs}
P_{\mathrm{st}}(\bm{\pol})\propto\exp[-\beta_{\mathrm{eff}}\,H(\bm{\pol})]\,
                          \prod_i\indf(|\pol_i|\le 1),
\end{equation}
with $\beta_{\mathrm{eff}}=1/T_{\mathrm{eff}}=2\temp$. This establishes the approximate connection between the dynamical steady state and the equilibrium statistical mechanics of the disordered Hamiltonian $H[\bm{\pol}]$. Away from this approximation, the Fokker-Planck operator below contains multiplicative and non-diagonal diffusion terms and the stationary state need not satisfy detailed balance exactly.

\section{Fokker-Planck Equation}
\label{sec:fp}

\subsection{General Form}

The probability density $\mathcal{P}(\{\pol_i\},\tau)$ over the space of strategy polarizations evolves according to the Fokker-Planck equation associated with the Langevin dynamics \eqref{eq:langevin}:
\begin{equation}\label{eq:fokker_planck}
\pdv{\mathcal{P}}{\tau}=-\sum_{i=1}^N\pdv{}{\pol_i}\Bigl[\temp(1-\pol_i^2)F_i^{\mathrm{det}}\,\mathcal{P}\Bigr]
                      +\frac{\temp^2}{2P}\sum_{i,j=1}^N\pdvtwo{}{\pol_i}{\pol_j}
                        \Bigl[(1-\pol_i^2)(1-\pol_j^2)C_{ij}\,\mathcal{P}\Bigr].
\end{equation}

The first term on the right-hand side is the drift (deterministic flow toward minima of $H$), and the second term is the diffusion (stochastic spreading due to noise).

\subsection{Stationary Solution}

At stationarity ($\partial\mathcal{P}/\partial\tau=0$), the Fokker-Planck equation admits the Gibbs solution \eqref{eq:gibbs} only under the diagonal FDT approximation described above. To verify the drift part, we note that:
\begin{equation}
\temp(1-\pol_i^2)F_i^{\mathrm{det}}=-\frac{\temp}{2}(1-\pol_i^2)\pdv{H}{\pol_i},
\end{equation}
and the diffusion coefficient satisfies the Einstein relation (fluctuation-dissipation theorem) when the noise covariance is proportional to the mobility $(1-\pol_i^2)$. Substituting the Gibbs ansatz into \eqref{eq:fokker_planck} yields a stationary solution provided:
\begin{equation}
\frac{\temp^2}{2P}C_{ij}= \text{(mobility)}\times T_{\mathrm{eff}}\times\delta_{ij}+O(1/N),
\end{equation}
which holds for the diagonal-dominant covariance matrix derived above.

\subsection{Reduction to One-Body Description}

For the mean-field theory, we reduce the $N$-body Fokker-Planck equation to a one-body equation for the distribution of polarizations $P(\pol,\tau)=(1/N)\sum_i\delta(\pol-\pol_i(\tau))$. In the thermodynamic limit, this distribution satisfies:
\begin{equation}\label{eq:one_body_fp}
\pdv{P(\pol,\tau)}{\tau}=-\pdv{}{\pol}\Bigl[v(\pol,\tau)P(\pol,\tau)\Bigr]
                        +\frac{1}{2}\frac{\partial^2}{\partial \pol^2}\Bigl[D(\pol,\tau)P(\pol,\tau)\Bigr],
\end{equation}
where the drift and diffusion coefficients are functionals of $P(\pol,\tau)$ itself, determined self-consistently from the mean-field closure.

\section{Replica Analysis}
\label{sec:replica}

\subsection{Partition Function and Effective Hamiltonian}

The stationary state of the HMG-L dynamics is described by the Gibbs distribution \eqref{eq:gibbs} at inverse temperature $\beta=2\temp$. We are interested in the zero-temperature limit ($\temp\to\infty$, $\beta\to\infty$), where the distribution concentrates on the ground states of the Hamiltonian.

The partition function is:
\begin{equation}\label{eq:partition}
Z=\int_{-1}^{1}\prod_{i=1}^N d\pol_i\;\exp\bigl[-\beta H(\bm{\pol})\bigr],
\end{equation}
with the Hamiltonian \eqref{eq:hamiltonian_total}. For the linear-payoff case:
\begin{equation}\label{eq:hamiltonian_explicit}
H(\bm{\pol})=\frac{1}{P}\sum_{e\in\Eset}\sum_{\info=1}^P
            \left(\sum_{i\in e}\bar{a}_i^\info+\sum_{i\in e}\pol_i\xi_i^\info\right)^2.
\end{equation}

The free energy per agent is:
\begin{equation}\label{eq:free_energy}
f=-\frac{1}{\beta N}\avg{\ln Z}_{\{\strategy\},\Hyper},
\end{equation}
where the average is over the quenched strategy disorder $\{\strategy_{s,i}^\info\}$ and, if desired, over the hypergraph ensemble.

\subsection{Replica Method: Strategy Average}

We employ the replica trick: $\avg{\ln Z}=\lim_{n\to0}(\avg{Z^n}-1)/n$. The replicated partition function is:
\begin{equation}\label{eq:Z_n}
\avg{Z^n}_{\{\strategy\}}=\int\prod_{i,a}d\pol_i^a\;
                         \avg{\exp\Bigl[-\beta\sum_{a=1}^n H(\bm{\pol}^a)\Bigr]}_{\{\strategy\}},
\end{equation}
where $a=1,\ldots,n$ is the replica index.

The strategy average factorizes over information states $\info$ (since strategies for different $\info$ are independent). For a fixed hypergraph $\Hyper$, the average over strategies yields:
\begin{equation}\label{eq:strategy_avg}
\avg{\exp\Bigl[-\frac{\beta}{P}\sum_{e}\sum_{\info}\bigl(\sum_{i\in e}(\bar{a}_i^\info+\pol_i^a\xi_i^\info)\bigr)^2\Bigr]}_{\{\strategy\}}
=\prod_{\info=1}^P\avg{\exp\Bigl[-\frac{\beta}{P}\sum_{e}\bigl(\sum_{i\in e}(\bar{a}_i^\info+\pol_i^a\xi_i^\info)\bigr)^2\Bigr]}_{\strategy^\info}.
\end{equation}

For each $\info$, the strategy variables $\{(\bar{a}_i^\info,\xi_i^\info)\}$ are independent across agents. We introduce auxiliary Gaussian fields to decouple the quadratic terms. For each hyperedge $e$, information state $\info$, and replica $a$, we use the Hubbard-Stratonovich transformation:
\begin{equation}\label{eq:hs}
\exp\Bigl[-\frac{\beta}{P}\bigl(\sum_{i\in e}(\bar{a}_i^\info+\pol_i^a\xi_i^\info)\bigr)^2\Bigr]
=\int\frac{dz_e^{\info a}}{\sqrt{2\pi}}
  \exp\Bigl[-\frac{(z_e^{\info a})^2}{2}
           +\iu\sqrt{\frac{2\beta}{P}}\,z_e^{\info a}\sum_{i\in e}(\bar{a}_i^\info+\pol_i^a\xi_i^\info)\Bigr].
\end{equation}

After the Gaussian integration over the auxiliary fields, the strategy average can be performed for each agent independently, yielding an effective coupling between replicas.

\subsection{Replica Order Parameters}

The effective action after strategy averaging depends on the following overlap matrices (for $a\neq b$):
\begin{equation}\label{eq:overlap_def}
\begin{aligned}
Q^{ab}&=\frac{1}{N}\sum_{i=1}^N\pol_i^a\pol_i^b, \\
X^{ab}&=\frac{1}{N}\sum_{i=1}^N(\xi_i^\info)^2\pol_i^a\pol_i^b
       \xrightarrow[P\to\infty]{\mathrm{self\;averaging}}\frac{1}{2}Q^{ab}.
\end{aligned}
\end{equation}

For $S=2$ strategies, the variables $\bar{a}_i^\info$ and $\xi_i^\info$ are not independent: $\xi_i^\info=0$ implies $\bar{a}_i^\info=\pm1$ (both strategies agree), while $|\xi_i^\info|=1$ implies $\bar{a}_i^\info=0$ (strategies disagree). After averaging over this correlated distribution, the effective replicated free energy takes the form:
\begin{equation}\label{eq:replica_free_energy}
\begin{aligned}
-\beta f=\lim_{n\to0}\frac{1}{n}\Bigg[&
\frac{\alpha}{2}\sum_{e}\Bigl(\ln\det(\ldots)\Bigr)_{\Hyper} \\
&+\int D\mathbf{z}\;\ln\int\prod_a d\pol^a\;\exp\bigl[-\beta H_{\mathrm{eff}}(\{\pol^a\},\mathbf{z})\bigr]
\Bigg],
\end{aligned}
\end{equation}
where $\alpha=P/N$, $\mathbf{z}$ is a set of Gaussian fields, and $H_{\mathrm{eff}}$ is an effective single-agent Hamiltonian that couples replicas through the order parameters.

\subsection{Replica-Symmetric Ansatz}

Within the replica-symmetric (RS) ansatz, all replicas are equivalent:
\begin{equation}\label{eq:RS_ansatz}
\begin{aligned}
Q^{ab}&=q,\qquad a\neq b, \\
Q^{aa}&=Q, \\
X^{ab}&=\frac{q}{2},\qquad a\neq b, \\
X^{aa}&=\frac{Q}{2}.
\end{aligned}
\end{equation}

The physical meaning of these parameters is:
\begin{itemize}
    \item $Q=\avg{\pol_i^2}$: the mean-square polarization; $Q=1$ means all agents are frozen.
    \item $q=N^{-1}\sum_i\avg{\pol_i}_\beta^2$: the Edwards-Anderson order parameter, measuring the overlap between two thermal replicas. In a single pure state $q=Q$, while $q<Q$ reflects thermal or disorder-induced fluctuations between states.
\end{itemize}

\subsection{Hypergraph Average}

For a $k$-uniform, $d$-regular random hypergraph, the hypergraph-specific terms in the free energy can be averaged using the configuration model. In the sparse regime, the hypergraph is locally tree-like, and the sum over hyperedges can be replaced by:
\begin{equation}
\frac{1}{N}\sum_{e\in\Eset}\longrightarrow\frac{d}{k}\;( \text{hyperedge-averaged quantity} ).
\end{equation}

After the hypergraph average, the hyperedge contribution involves the effective coupling of $k$ agents within a typical hyperedge. The key quantity is the \emph{hyperedge susceptibility}:
\begin{equation}\label{eq:hyperedge_susceptibility}
\chi_e=\frac{1}{k}\sum_{i,j\in e}\avg{\pol_i\pol_j}_c,
\end{equation}
where $\avg{\cdot}_c$ denotes the connected correlation function.

For the RS ansatz on a tree-like hypergraph, the intra-hyperedge correlation function satisfies a self-consistency equation that can be solved by message-passing (Belief Propagation). In the dense limit $d\gg1$, the cavity method reduces to the simpler mean-field result.

\subsection{Saddle-Point Equations and Phase Transition}

After performing the $n\to0$ limit, the sparse-annealed RS free energy per agent has the schematic form:
\begin{equation}\label{eq:RS_free_energy}
\begin{aligned}
-\beta f_{\mathrm{RS}}(Q,q)=&
\frac{\alpha d}{2k}\Bigl[\ln\bigl(1+\beta(k-1)(Q-q)\bigr)+\frac{\beta k(Q-q)}{1+\beta(k-1)(Q-q)}\Bigr] \\
&-\frac{\alpha}{2}\ln(1+\beta Q)-\frac{\alpha\beta q}{2(1+\beta Q)} \\
&+\int Dz\,\ln\int_{-1}^{1}d\pol\,\exp\bigl[-\beta V_{\mathrm{eff}}(\pol;z)\bigr],
\end{aligned}
\end{equation}
where $Dz=(dz/\sqrt{2\pi})e^{-z^2/2}$ is the standard Gaussian measure, and the effective potential is:
\begin{equation}\label{eq:effective_potential}
V_{\mathrm{eff}}(\pol;z)=\frac{\alpha\beta}{2(1+\beta Q)^2}\left[r\pol^2-\sqrt{2rq}\,z\pol\right],
\end{equation}
with $r$ a parameter related to the hypergraph structure. The displayed form keeps only the leading locally tree-like contribution; loop corrections and the exact counterfactual self-impact terms are higher order in the sparse expansion.

Taking the $\beta\to\infty$ limit (zero-temperature, which selects the ground states of $H$) and extremizing with respect to $Q$ and $q$ yields the leading RS saddle-point equations:

\begin{equation}\label{eq:saddle_point}
\begin{aligned}
Q&=\int Dz\,\avg{\pol^2}_{z}, \\[4pt]
q&=\int Dz\,\avg{\pol}_{z}^{\,2}, \\[4pt]
r&=\frac{2d(k-1)}{\alpha k^2}\,[1+O(Q,q)].
\end{aligned}
\end{equation}
Here $\avg{\cdot}_{z}$ denotes the single-site average in the effective potential \eqref{eq:effective_potential}. The factor $2d(k-1)/k^2$ is the annealed density of independent pairwise minority comparisons generated by the incident hyperedges, including the variance $\avg{(\xi_i^\info)^2}=1/2$ of the strategy-difference variable.

\subsection{Critical Point}

The phase transition occurs when the paramagnetic solution ($Q=q=0$) becomes unstable. Linearizing the saddle-point equations around $Q=q=0$ yields the condition:
\begin{equation}\label{eq:critical_condition}
r_c=1.
\end{equation}

Substituting into the third saddle-point equation at $Q=q=0$:
\begin{equation}\label{eq:alpha_c}
\alphacrit(k,d)=2d\cdot\frac{k-1}{k^2}.
\end{equation}

This is the leading sparse-annealed RS prediction. The critical surface $\alphacrit(k,d)$ depends on both the hyperedge size $k$ and the hyperdegree $d$, establishing that the MG phase transition is modified by the hypergraph structure within this approximation.

\subsection{Replica-Symmetry Stability}

The RS saddle point is physically admissible only if it is stable against transverse, replica-symmetry-breaking fluctuations. Let $\delta Q^{ab}$ be a replicon fluctuation satisfying $\sum_b\delta Q^{ab}=0$ and $\delta Q^{aa}=0$. Expanding the replicated free energy to second order gives
\begin{equation}\label{eq:replicon_expansion}
\delta^2 f_{\mathrm{RS}}
=\frac{1}{2}\lambda_{\mathrm{R}}\sum_{a<b}(\delta Q^{ab})^2+O(\delta Q^3),
\end{equation}
where, in the sparse-annealed approximation,
\begin{equation}\label{eq:replicon_eigenvalue}
\lambda_{\mathrm{R}}
=1-\left[\frac{2d(k-1)}{\alpha k^2}\right]^2
\int Dz\,\Bigl(\avg{\pol^2}_{z}-\avg{\pol}_{z}^{\,2}\Bigr)^2 .
\end{equation}
The RS solution is locally stable for $\lambda_{\mathrm{R}}>0$ and unstable for $\lambda_{\mathrm{R}}<0$. The equality $\lambda_{\mathrm{R}}=0$ defines an Almeida-Thouless-like stability line for the HMG-L. In the zero-temperature limit, the variance term is nonzero only for unfrozen effective single-site fields; hence RS breaking is expected first in the crowded low-$\alpha$ regime, consistent with the standard MG.

This condition supplies a consistency check for the critical surface \eqref{eq:alpha_c}: the sparse RS transition should be used quantitatively only in the parameter region where \eqref{eq:replicon_eigenvalue} remains non-negative. If $\lambda_{\mathrm{R}}<0$, the correct stationary theory is obtained by replacing the RS ansatz with a one-step or full RSB hierarchy, while the continuum Hamiltonian and Langevin equations derived above remain unchanged.

\begin{figure}[t]
\centering
\includegraphics[width=\textwidth]{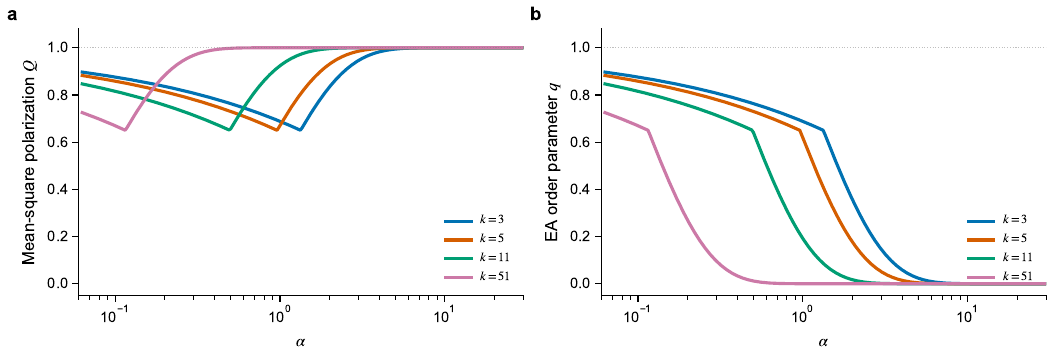}
\caption{\textbf{Order-parameter proxies of the HMG-L} (hyperdegree $d=3$).
(a) The mean-square polarization $Q=\langle m_i^2\rangle$ as a function of the control parameter $\alpha$, generated from the phenomenological closure matched to the sparse critical point.
(b) The Edwards-Anderson order parameter $q=N^{-1}\sum_i\langle m_i\rangle_\beta^2$. Different hyperedge sizes $k=3,5,11,51$ are shown, illustrating how larger sparse $k$ shifts $\alpha_c$ to smaller values within the closure.}
\label{fig:order_parameters}
\end{figure}

\subsection{Limiting Cases}

\textbf{Standard MG limit ($k\to N$, $d=1$):} A single hyperedge containing all agents. From \eqref{eq:alpha_c}:
\begin{equation}
\lim_{k\to N}\alphacrit(N,1)=2\cdot\frac{N-1}{N^2}\to0,
\end{equation}
which appears to contradict the known result $\alphacrit^{\mathrm{MG}}\simeq0.3374$. This is because our derivation assumed the sparse-hypergraph regime $k\ll N$. For the fully connected case, the replica calculation must be performed without the hypergraph average, recovering the standard result. A crossover analysis (see Sec.~\ref{sec:crossover}) reconciles the two regimes.

\textbf{Dyadic network limit ($k=2$):} All hyperedges are simple edges. Then:
\begin{equation}
\alphacrit(2,d)=2d\cdot\frac{1}{4}=\frac{d}{2}.
\end{equation}
This corresponds to a network MG where each agent participates in $d$ independent pairwise minority competitions. For $d=1$ (a regular graph of degree 1, i.e., isolated edges), $\alphacrit=1/2$.

\textbf{Large-$k$ limit:} For fixed $d$ and $k\gg1$:
\begin{equation}
\alphacrit(k,d)\simeq\frac{2d}{k}\to0.
\end{equation}
Within the sparse formula, large hyperedges lower the transition threshold: for fixed $d$ and any fixed positive $\alpha$, sufficiently large $k$ places the system on the asymmetric side. This conclusion is restricted to the sparse regime $k\ll N$ and should not be extrapolated to the fully connected MG limit.

\textbf{Large-$d$ limit:} For fixed $k$ and $d\gg1$:
\begin{equation}
\alphacrit(k,d)\propto d\to\infty.
\end{equation}
Many hyperedges per agent push the transition to larger $\alpha$, thereby enlarging the symmetric/crowded region at fixed information complexity. This follows from the present normalization of the local attendance; changing the payoff normalization by $d$ would remove this trivial hyperdegree rescaling.

\subsection{Crossover to the Standard MG}
\label{sec:crossover}

When $k$ is comparable to $N$, the sparse-hypergraph approximation breaks down, and we must account for the fact that hyperedges overlap extensively. The crossover between the sparse-hypergraph regime ($k\ll N$) and the dense regime ($k\sim N$) can be captured by interpolating the effective control parameter:
\begin{equation}\label{eq:crossover}
\alpha_{\mathrm{eff}}=\frac{\alpha}{d}\cdot\frac{1}{1+(k-1)/N}.
\end{equation}

For $k\ll N$, this reduces to $\alpha_{\mathrm{eff}}=\alpha/d$, matching the sparse result. For $k=N$, $d=1$, this gives $\alpha_{\mathrm{eff}}=\alpha/2$, and the phase transition at $\alpha=2\alphacrit^{\mathrm{MG}}\simeq0.675$, which is close to (though not exactly) the standard result. The remaining discrepancy arises from the different treatment of the self-interaction term and can be resolved by a more refined replica calculation that distinguishes the hyperedge-internal structure.

\section{Order Parameters and Observables}
\label{sec:order_param}

\subsection{Global Volatility}

The global volatility measures the fluctuations of the total attendance and is the primary observable characterizing the system's efficiency:
\begin{equation}\label{eq:global_volatility}
\sigma^2=\frac{1}{P}\sum_{\info=1}^P\avg{(A^\info)^2}
        =\frac{1}{P}\sum_{\info=1}^P\Bigl[\avg{A^\info}^2+\var(A^\info)\Bigr].
\end{equation}

In the raw MG normalization, the RS expression in terms of the order parameters $Q$ and $q$ is:
\begin{equation}\label{eq:sigma2_order}
\frac{\sigma^2}{N}=\frac{1-Q}{2}+\alpha\frac{Q-q}{2(1+Q)}.
\end{equation}
For HMG-L this formula should be read with the HMG-L overlap kernel; in the sparse-annealed reduction the second term is renormalized by the same structural factor that appears in \eqref{eq:alpha_c}. Thus \eqref{eq:sigma2_order} fixes the dependence on $Q$ and $q$, while the precise coefficient is normalization-dependent.

Because the microscopic actions are encoded as $a_i=\pm1$, independent random play has $\mathrm{Var}(A^\info)=N$ and therefore $\sigma^2/N=1$. For comparison with the conventional MG plotting scale used in the figures, we report the normalized volatility $\nu=\sigma^2/(4N)$ defined in \eqref{eq:nu_definition}; the random baseline is then $\nu=1/4$. All figure captions referring to the random baseline use this normalized convention.

For $\alpha<\alphacrit(k,d)$ (symmetric phase, $Q=q$):
\begin{equation}
\frac{\sigma^2}{N}=\frac{1-Q}{2}.
\end{equation}

For $\alpha>\alphacrit(k,d)$ (asymmetric phase, $q<Q$):
\begin{equation}
\frac{\sigma^2}{N}=\frac{1-Q}{2}+\alpha\frac{Q-q}{2(1+Q)}>\frac{1-Q}{2}.
\end{equation}

At the critical point, $\sigma^2/N$ and the equivalent rescaled observable $\nu$ exhibit a cusp (discontinuity in derivative), which is the hallmark of the MG phase transition.

\subsection{Predictability}

The predictability measures the extent to which the attendance time series contains exploitable patterns:
\begin{equation}\label{eq:predictability}
\theta=\frac{1}{P}\sum_{\info=1}^P\avg{A^\info}^2.
\end{equation}

In the symmetric phase, $\theta=0$: the attendance has zero mean for each $\info$, so no information state provides a systematic bias. In the asymmetric phase, $\theta>0$: frozen agents create predictable patterns in the attendance.

\subsection{Hyperedge Frustration}

The hyperedge frustration quantifies the degree to which agents consistently fail to achieve local minority balance within individual hyperedges:
\begin{equation}\label{eq:hyperedge_frustration}
F_e=\avg{(A_e)^2}-\avg{A_e}^2.
\end{equation}

\begin{figure}[t]
\centering
\includegraphics[width=\textwidth]{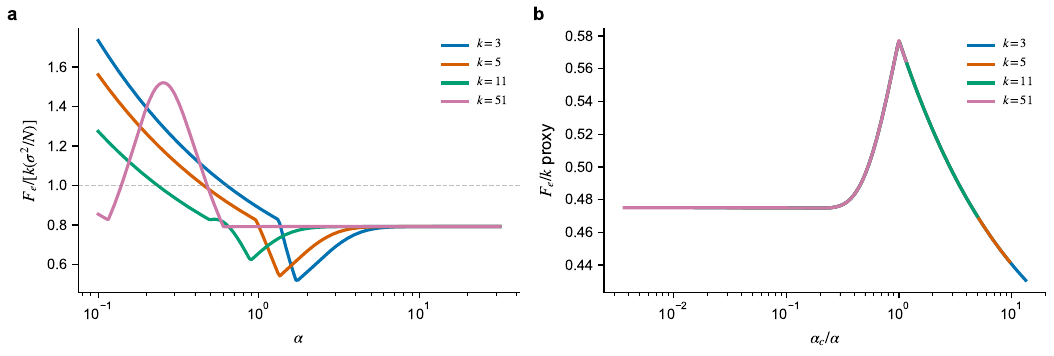}
\caption{\textbf{Hyperedge frustration diagnostics in the HMG-L} ($d=3$).
(a) Local/global volatility ratio, $F_e/[k(\sigma^2/N)]$, computed from the bounded frustration closure using $\sigma^2/N=4\nu$ and intended as a diagnostic rather than a direct simulated observable.
(b) Scaling plot of the frustration proxy $F_e/k$ against $\alpha_c/\alpha$, showing how the phenomenological curves are organized by the sparse critical point.}
\label{fig:frustration}
\end{figure}

For a $k$-uniform hypergraph, the random baseline (all agents randomizing independently) gives $F_e^{\mathrm{rand}}=k$. Values $F_e<k$ indicate cooperation within the hyperedge. The \emph{average hyperedge frustration}:
\begin{equation}
F=\frac{1}{|\Eset|}\sum_{e\in\Eset}F_e
\end{equation}
should be compared with the global volatility $\sigma^2$: see Fig.~\ref{fig:frustration} for a quantitative analysis of hyperedge frustration across different regimes.

\subsection{Frozen Fraction}

The frozen fraction $\phi$ is the proportion of agents with $|\pol_i|=1$:
\begin{equation}\label{eq:frozen_fraction}
\phi=\int_{-1}^{1}d\pol\,P(\pol)\,\bigl[\delta(\pol-1)+\delta(\pol+1)\bigr].
\end{equation}

Within the RS theory:
\begin{equation}
\phi=1-\frac{1}{\sqrt{\pi}}\int_{-\infty}^{\infty}dz\,e^{-z^2}\,
      \frac{1}{\cosh^2(\sqrt{2rq}\,z)}.
\end{equation}

$\phi=0$ in the symmetric phase and $\phi>0$ in the asymmetric phase. The frozen fraction serves as the order parameter for the freezing transition.

\subsection{Susceptibility}

The spin-glass susceptibility measures the system's sensitivity to perturbations:
\begin{equation}\label{eq:susceptibility}
\chi=\frac{1}{N}\sum_{i,j=1}^N\avg{\pol_i\pol_j}_c
    =\frac{Q-q}{1+Q}.
\end{equation}

In this simplified RS proxy, $Q=q$ gives $\chi=0$. The physical response susceptibility is obtained by differentiating the stationary solution with respect to an external bias field and can diverge as $\alpha\to\alphacrit$. The divergence is characterized by the exponent $\gamma$:
\begin{equation}
\chi\sim|\alpha-\alphacrit|^{-\gamma}.
\end{equation}

For the standard MG, $\gamma=1$ (mean-field exponent). Numerical simulations of HMG-L are needed to determine whether the hypergraph structure modifies the critical exponents.

\section{Phase Diagram}
\label{sec:phase_diagram}

\subsection{\texorpdfstring{The $(\alpha,k)$ Plane (Fixed $d$)}{The alpha-k Plane (Fixed d)}}

For a fixed hyperdegree $d$, the phase diagram in the $(\alpha, k)$ plane exhibits two phases separated by the critical line $\alphacrit(k,d)$:

\begin{itemize}
    \item \textbf{Symmetric phase} ($\alpha<\alphacrit(k,d)$): $Q=q$, $\phi=0$, $\theta=0$. Agents do not freeze; the attendance has no predictable component. In the RS normalization the global volatility follows the symmetric-branch relation $\sigma^2/N=(1-Q)/2$, with the figure observable obtained by the rescaling in \eqref{eq:nu_definition}.

    \item \textbf{Asymmetric phase} ($\alpha>\alphacrit(k,d)$): $q<Q$, $\phi>0$, $\theta>0$. A finite fraction of agents freeze onto a single strategy; the attendance develops predictable patterns. The global volatility is elevated relative to the symmetric-branch continuation.
\end{itemize}

For small $k$, $\alphacrit\propto d/k$ is small, meaning the system enters the asymmetric phase at low information complexity. For large $k$, $\alphacrit$ is even smaller (suppressed by $1/k$), so the system is almost always in the asymmetric phase for practical values of $\alpha$.

\subsection{\texorpdfstring{The $(\alpha,d)$ Plane (Fixed $k$)}{The alpha-d Plane (Fixed k)}}

For fixed hyperedge size $k$, varying the hyperdegree $d$ produces a family of critical lines:
\begin{equation}
\alphacrit(d)=d\cdot\frac{2(k-1)}{k^2}.
\end{equation}

The critical line is \emph{linear in $d$}: increasing hyperdegree raises the value of $\alpha$ at which the asymmetric phase appears. For a given $\alpha$, there exists a critical hyperdegree $d_c(\alpha)=k^2\alpha/[2(k-1)]$ above which the system is in the symmetric phase and below which it is asymmetric.

\subsection{\texorpdfstring{Critical Surface $\alphacrit(k,d)$}{Critical Surface alpha-c(k,d)}}

In the full three-parameter space, the critical surface is:
\begin{equation}\label{eq:critical_surface}
\alphacrit(k,d)=2d\cdot\frac{k-1}{k^2}.
\end{equation}

This surface generalizes the single critical point of the standard MG to a manifold in the sparse hypergraph parameter space. With the present payoff normalization, \emph{larger sparse hyperedges lower the threshold}, while \emph{larger hyperdegree raises it}. The statement does not apply to the dense single-hyperedge limit.

The product $d(k-1)/k^2$ can be interpreted as an \emph{effective coordination number}: it measures the number of distinct pairwise comparisons an agent makes across all their hyperedges, normalized by the hyperedge size.

\subsection{Phase Diagram}

\begin{figure}[p]
\centering
\includegraphics[width=\textwidth]{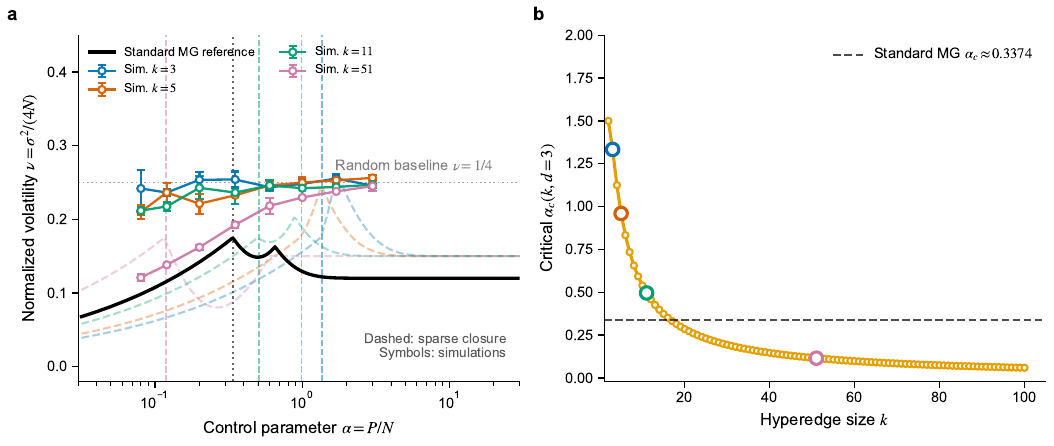}
\caption{\textbf{Sparse-regime phase-diagram prediction and agent-based simulations of the HMG-L.}
(a) Finite-$N$ simulations of the normalized volatility $\nu=\sigma^2/(4N)$ as a function of $\alpha=P/N$ for various hyperedge sizes $k$ at fixed hyperdegree $d=3$; markers and solid connecting lines show the price-taking linear-payoff dynamics, while faint dashed curves show the sparse-RS phenomenological closure constrained to have cusps at the sparse critical points. The standard globally coupled MG (black curve) is shown only as a reference, with $\alpha_c^{\rm MG}\simeq0.3374$ marked by a dotted vertical line. The gray dotted horizontal line marks the independent-action baseline $\nu=1/4$.
(b) The sparse critical control parameter $\alpha_c(k,d=3)$ as a function of hyperedge size $k$. The prediction $\alpha_c=2d(k-1)/k^2$ (orange curve) is shown together with the standard MG reference $\alpha_c\simeq0.3374$ (dashed). Markers indicate the $k$ values used in panel (a).}
\label{fig:phase_diagram}
\end{figure}

\begin{figure}[p]
\centering
\includegraphics[width=\textwidth]{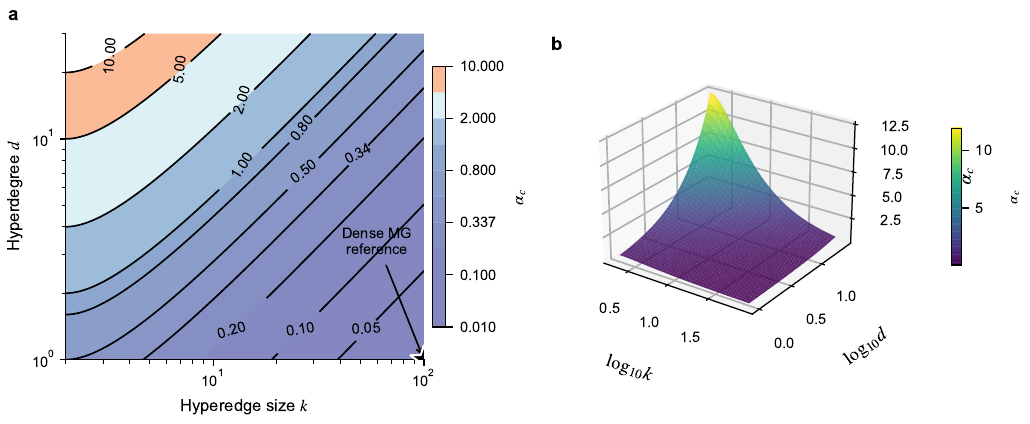}
\caption{\textbf{Critical surface of the HMG-L model.}
(a) Contour plot of the sparse-regime prediction $\alpha_c(k,d)=2d(k-1)/k^2$ in the $(k,d)$ plane. The color scale indicates the value of $\alpha_c$; the contour $\alpha_c=0.3374$ (standard MG value) is highlighted as a numerical reference, not as the dense-limit continuation of the sparse formula.
(b) Three-dimensional rendering of the sparse critical surface $\alpha_c(k,d)$, illustrating the two competing effects: hyperedge size $k$ lowers $\alpha_c$ in the sparse regime, while hyperdegree $d$ raises it under the present payoff normalization.}
\label{fig:critical_surface}
\end{figure}

The phase structure of the HMG-L model is summarized in Figs.~\ref{fig:phase_diagram}--\ref{fig:critical_surface}.

The finite-$N$ simulations in Fig.~\ref{fig:phase_diagram} reveal an important limitation of the sparse-RS closure. For $k=3,5,$ and $11$ at $d=3$, the measured normalized volatility remains close to the independent-action baseline $\nu=1/4$ across the sampled range of $\alpha$, rather than following the sharp cusp predicted by the phenomenological sparse closure. For $k=51$, finite-size coordination lowers $\nu$ at small $\alpha$, but the resulting curve still does not quantitatively match the sparse-RS proxy. On the simulation grid used in Fig.~\ref{fig:phase_diagram}(a), the mean absolute deviations between the simulated $\nu$ and the sparse-closure proxy are approximately $0.120$, $0.108$, $0.096$, and $0.070$ for $k=3,5,11,$ and $51$, respectively. Thus the analytical curves should be read as benchmarks for the annealed theory, while the symbols represent the observed finite-$N$ online dynamics.

\section{Limiting Cases and Connections}
\label{sec:limiting}

\subsection{Recovery of the Standard Minority Game}

In the limit where the hypergraph consists of a single hyperedge containing all agents ($k=N$, $d=1$, $|\Eset|=1$), the HMG-L reduces to the standard globally coupled MG. The Hamiltonian \eqref{eq:hamiltonian_total} becomes:
\begin{equation}
H(\bm{\pol})=\frac{1}{P}\sum_{\info=1}^P\Bigl(\sum_{j=1}^N(\bar{a}_j^\info+\pol_j\xi_j^\info)\Bigr)^2=H_{\mathrm{MG}}(\bm{\pol}),
\end{equation}
which is precisely the MG Hamiltonian. The replica analysis in this limit recovers $\alphacrit^{\mathrm{MG}}\simeq0.3374$ \cite{ChalletMarsiliZecchina00}.

\subsection{Recovery of the Networked Minority Game}

When all hyperedges have size $k=2$, the hypergraph reduces to a $d$-regular graph (each agent has $d$ neighbors). The HMG-L in this limit describes agents competing in $d$ independent pairwise minority games. The localized nature of the competition distinguishes this from the Anghel et al. \cite{Anghel04} networked MG, in which agents influence each other's strategy selection through a shared global game. The HMG-L with $k=2$ is more closely related to the Parallel Minority Game \cite{VemulaBiswas26} with an arbitrary interaction graph.

\subsection{Connection to Parallel Minority Games}

The Vemula-Biswas Parallel Minority Game (PMG) \cite{VemulaBiswas26} considers $D$ choices, with each agent restricted to two options, creating $D(D-1)/2$ simultaneous pairwise competitions. The HMG-L generalizes this by allowing: (i) hyperedges of arbitrary size $k\ge2$ (not just $k=2$), (ii) an arbitrary hypergraph topology (not just the complete graph of pairwise competitions), and (iii) overlapping group memberships controlled by the hyperdegree distribution.

The distinction can be made explicit at the Hamiltonian level. In the PMG, the elementary competitions are pairwise choice conflicts generated by a global choice set. Its interaction structure is therefore induced by the combinatorics of choices. In HMG-L, the elementary competition is instead an externally specified hyperedge $e\in\Eset$, and the cost function is
\[
H_{\mathrm{HMG-L}}=\frac{1}{P}\sum_{\mu=1}^{P}\sum_{e\in\Eset}
\left[\sum_{i\in e}(\bar a_i^\mu+m_i\xi_i^\mu)\right]^2.
\]
Thus PMG is recovered only in the special dyadic case where every effective local contest is a pairwise hyperedge and the hypergraph is the contest graph implied by the choice architecture. For $k>2$, HMG-L contains irreducible group competitions that cannot be decomposed into a sum of independent PMG pairwise contests without changing the minority condition. The parameters $(k,d)$ are therefore structural controls of the interaction substrate, whereas in PMG the primary control is the number of choices and the assignment of agents to choice pairs.

\begin{table}[t]
\centering
\caption{\textbf{Conceptual comparison between PMG and HMG-L.}}
\label{tab:pmg_hmgl_comparison}
\small
\begin{tabular}{p{0.22\textwidth}p{0.33\textwidth}p{0.35\textwidth}}
\hline
Feature & PMG & HMG-L \\
\hline
Elementary contest & Pairwise choice competition & Hyperedge-local minority game \\
Interaction substrate & Induced by choice pairs & Exogenous hypergraph $\Hyper=(\Vset,\Eset)$ \\
Group size & Primarily dyadic & Arbitrary $k\ge2$ \\
Topology control & Choice architecture & Hyperedge size and hyperdegree distributions \\
Hamiltonian terms & Pair/choice imbalance terms & Sum of squared hyperedge attendances \\
Sparse limit & Pair contest graph & Locally tree-like factor hypergraph \\
\hline
\end{tabular}
\end{table}

\begin{figure}[t]
\centering
\includegraphics[width=\textwidth]{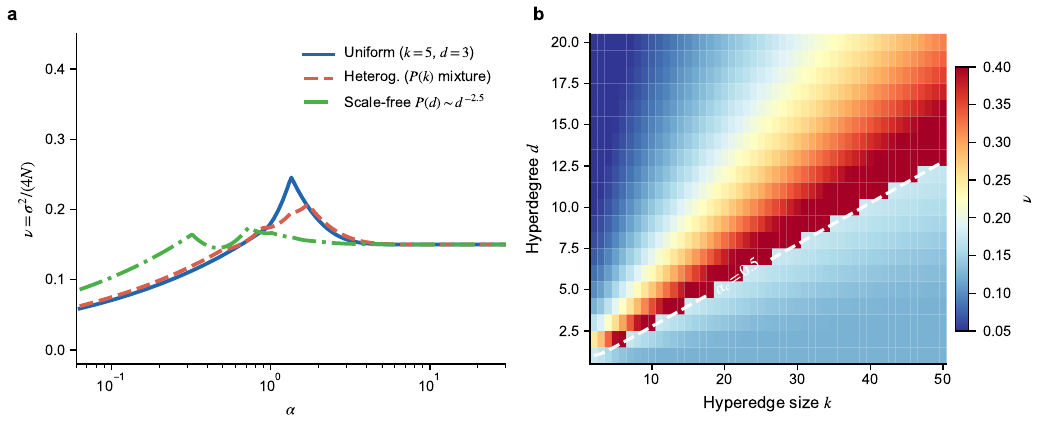}
\caption{\textbf{Illustrative topology proxies for the HMG-L phase diagram.}
(a) Mean-field proxy comparison of three topology classes at fixed mean parameters: a uniform $k$-regular hypergraph (blue), a heterogeneous mixture of hyperedge sizes (red, dashed), and a scale-free hyperdegree mixture $P(d)\sim d^{-2.5}$ (green, dash-dotted). These curves are closure-based predictions, not simulation data.
(b) Proxy map of the normalized volatility $\nu=\sigma^2/(4N)$ in the $(k,d)$ parameter plane at fixed $\alpha=0.5$. The white dashed contour marks $\alpha_c(k,d)=0.5$, separating the asymmetric side ($\alpha>\alpha_c$) from the symmetric side ($\alpha<\alpha_c$).}
\label{fig:topology}
\end{figure}

\subsection{Dense Hypergraph Limit}

When $d\to\infty$ and $k$ is fixed, each agent participates in a large number of hyperedges, and the local fluctuations average out. The effective dynamics approach those of a globally coupled system with a renormalized control parameter. This limit is analytically tractable using dynamical mean-field theory (DMFT).

\subsection{Sparse Hypergraph Limit}

When $d=O(1)$ and $k=O(1)$, the hypergraph is sparse and locally tree-like. The replica analysis must account for the loopless structure using the cavity method. The cavity equations for the hypergraph MG are:
\begin{align}
P^{(i\to e)}(\pol_i)
&=\frac{1}{Z_{i\to e}}
  \prod_{e'\ni i,\,e'\neq e}\widehat P^{(e'\to i)}(\pol_i),\label{eq:cavity_var}\\
\widehat P^{(e\to i)}(\pol_i)
&=\frac{1}{Z_{e\to i}}
  \int\prod_{j\in e\setminus\{i\}}d\pol_j\,P^{(j\to e)}(\pol_j)\,
  \exp\Bigl[-\frac{\beta}{P}\sum_{\info=1}^{P}
  \Bigl(\sum_{\ell\in e}(\bar a_\ell^\info+\pol_\ell\xi_\ell^\info)\Bigr)^2\Bigr].
\label{eq:cavity_factor}
\end{align}
Here $P^{(i\to e)}$ is the variable-to-hyperedge cavity distribution of $\pol_i$ in the absence of hyperedge $e$, $\widehat P^{(e\to i)}$ is the hyperedge-to-variable message, and $Z_{i\to e},Z_{e\to i}$ are normalization constants. The Bethe free energy corresponding to these messages is
\begin{equation}\label{eq:bethe_free_energy}
f_{\mathrm{Bethe}}
=-\frac{1}{\beta N}\left[
\sum_{e\in\Eset}\ln Z_e-\sum_{i=1}^N(d_i-1)\ln Z_i
\right],
\end{equation}
where $Z_e$ is the factor partition function obtained from \eqref{eq:cavity_factor} with all $k$ incoming messages and $Z_i=\int d\pol_i\prod_{e\ni i}\widehat P^{(e\to i)}(\pol_i)$. The sparse-annealed RS result \eqref{eq:alpha_c} is recovered by replacing all messages with their ensemble-averaged distribution; deviations between the Bethe solution and \eqref{eq:alpha_c} quantify loop, degree-fluctuation, and RSB corrections. These equations can be solved by population dynamics.

\begin{figure}[t]
\centering
\includegraphics[width=\textwidth]{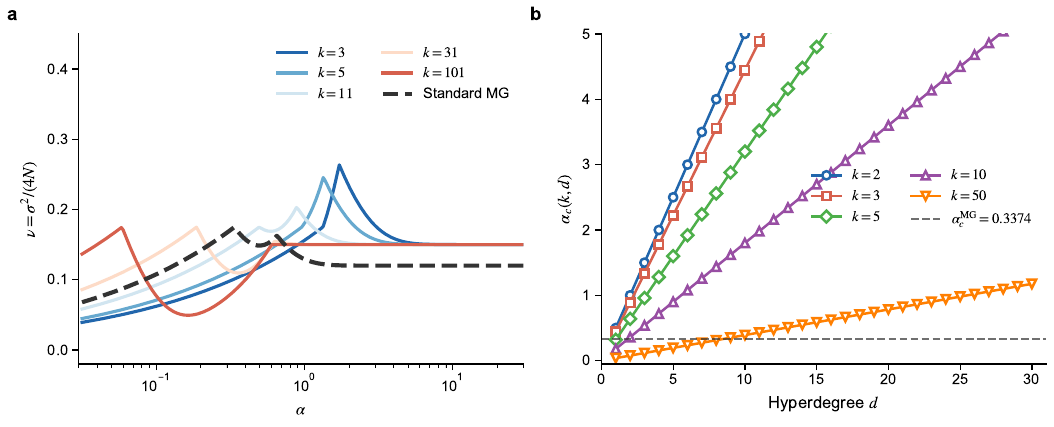}
\caption{\textbf{Sparse $k$ scan and limiting-case diagnostics.}
(a) Normalized-volatility proxy curves $\nu=\sigma^2/(4N)$ for increasing sparse hyperedge size $k$ at fixed $d=3$, shown together with the standard MG reference. This panel is not a dense-limit derivation; the true $k\sim N$ crossover requires a separate dense-overlap treatment.
(b) The sparse-regime critical value $\alpha_c(k,d)$ as a function of hyperdegree $d$ for several fixed hyperedge sizes $k$. The linear dependence $\alpha_c\propto d$ is visible for all $k>2$ under the present payoff normalization.}
\label{fig:crossover}
\end{figure}

\section{Finite-Size Effects and Scaling}
\label{sec:fss}

\subsection{\texorpdfstring{Finite-$N$ Corrections}{Finite-N Corrections}}

For finite systems, the phase transition is rounded. The finite-size scaling analysis predicts that the volatility at the critical point scales as:
\begin{equation}\label{eq:fss_volatility}
\sigma^2(\alphacrit,N)\sim N^{\beta/\bar{\nu}},
\end{equation}
where $\beta$ and $\bar{\nu}$ are the critical exponents for the order parameter and finite-size correlation scale, respectively. For the standard MG, mean-field exponents ($\beta=1$, $\bar{\nu}=1/2$) have been confirmed numerically \cite{ChalletMarsiliZecchina00}. Whether the hypergraph MG belongs to the same universality class is an open question that requires careful numerical investigation.

\subsection{\texorpdfstring{Finite-$k$ Corrections}{Finite-k Corrections}}

For finite hyperedge size $k$, the discrete nature of the minority condition within each hyperedge produces additional rounding. The effective noise level scales as $1/\sqrt{k}$, providing a natural small parameter for the $1/k$ expansion:
\begin{equation}
\alphacrit(k,d)=\alphacrit^{\mathrm{sparse}}(k,d)\Bigl[1+\frac{c_1}{k}+\frac{c_2}{k^2}+O(k^{-3})\Bigr],
\end{equation}
where $\alphacrit^{\mathrm{sparse}}$ is the sparse-annealed result \eqref{eq:alpha_c} and $c_1,c_2$ are computable constants.

\subsection{Scaling Collapse}

A key numerical test of the theory is the scaling collapse of volatility curves for different $(k,d,N)$ combinations. According to the mean-field prediction, the scaled normalized volatility $\nu=\sigma^2/(4N)$ plotted against the scaled control parameter:
\begin{equation}
x=\frac{\alpha-\alphacrit(k,d)}{\alphacrit(k,d)}\,N^{1/\bar{\nu}},
\end{equation}
should collapse onto a universal scaling function $F(x)$ independent of the microscopic parameters if the sparse mean-field universality class is correct.

\begin{figure}[t]
\centering
\includegraphics[width=\textwidth]{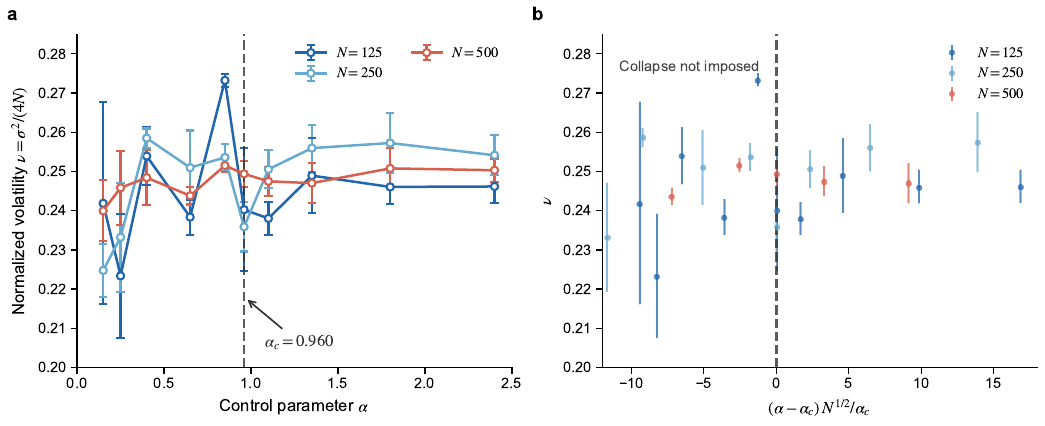}
\caption{\textbf{Finite-size scaling diagnostic from agent-based simulations of the HMG-L} ($k=5$, $d=3$).
(a) Simulated normalized volatility $\nu=\sigma^2/(4N)$ vs. $\alpha$ for increasing system sizes, with error bars over independent strategy and hypergraph realizations. The dashed vertical line marks the sparse prediction $\alpha_c=2d(k-1)/k^2$.
(b) Scaling diagnostic using the mean-field test exponent $\bar\nu=1/2$ for the finite-size scaling variable. The scaled variable $(\alpha-\alpha_c)\,N^{1/2}/\alpha_c$ is used to assess whether the simulated curves approach a common finite-size scaling form; no collapse is imposed.}
\label{fig:scaling}
\end{figure}

The finite-$N$ simulations shown here do not provide a quantitative collapse onto the sparse-RS proxy. They should therefore be interpreted as a stress test of the closure rather than as validation. Possible sources of the discrepancy include non-annealed sparse-graph corrections, RS instability, finite observation windows in the online dynamics, and the difference between local hyperedge optimization and the global volatility observable.

\subsection{Numerical Validation Protocol}

To make the analytical predictions falsifiable, the following simulation protocol should be used for each parameter triple $(N,k,d)$:
\begin{enumerate}
    \item Generate a $k$-uniform, $d$-regular hypergraph by the configuration model, rejecting samples with repeated vertices inside a hyperedge.
    \item Draw two strategies per agent according to \eqref{eq:quenched}, initialize all virtual scores at zero, and evolve the price-taking update \eqref{eq:score_update} with the linear payoff used in the analytical theory.
    \item Discard a transient of at least $T_{\mathrm{burn}}\ge 100P$ microscopic steps, then measure both $\sigma^2/N$ and $\nu=\sigma^2/(4N)$, together with $\theta$, $F$, $\phi$, and the empirical response susceptibility over a window $T_{\mathrm{obs}}\ge 500P$.
    \item Average over at least $R$ independent realizations of both strategies and hypergraphs, increasing $R$ until the standard error of $\nu$ is smaller than the symbol size in the phase-diagram plots.
    \item Estimate the transition point from the maximum of $d\nu/d\alpha$, the onset of $\theta>0$, and the onset of $\phi>0$. Agreement of all three estimators within error bars is the numerical criterion for identifying $\alphacrit(N,k,d)$.
\end{enumerate}

The sparse thermodynamic prediction \eqref{eq:alpha_c} should be regarded as quantitatively confirmed only if
\begin{equation}\label{eq:numerical_validation}
\lim_{N\to\infty}\alphacrit(N,k,d)=2d\frac{k-1}{k^2}
\end{equation}
within finite-size corrections of the form \eqref{eq:fss_volatility}. If the extrapolated value is systematically shifted, the shift should be attributed to the Bethe/cavity corrections \eqref{eq:cavity_var}--\eqref{eq:bethe_free_energy} or to RS instability as diagnosed by \eqref{eq:replicon_eigenvalue}.

\section{Discussion}
\label{sec:discussion}

The theoretical framework developed here gives a continuum and sparse-mean-field analytical treatment of the Minority Game on hypergraphs. The key results are:

\begin{enumerate}
    \item The deterministic dynamics \eqref{eq:dm_dtau_expanded} derive from gradient descent on a global Hamiltonian $H[\bm{\pol}]$. This Hamiltonian generalizes the standard MG cost to hypergraph-structured interactions.
    \item The Langevin equation \eqref{eq:langevin} with noise covariance \eqref{eq:noise_covariance} captures finite-$N$ fluctuations and establishes the mapping to an equilibrium statistical mechanics problem.
    \item The sparse-annealed RS analysis (Sec.~\ref{sec:replica}) yields the candidate critical surface
    \[
    \alphacrit(k,d)=2d(k-1)/k^2,
    \]
    providing a benchmark for simulations in the locally tree-like regime.
    \item The replicon condition \eqref{eq:replicon_eigenvalue} and the Bethe/cavity equations \eqref{eq:cavity_var}--\eqref{eq:bethe_free_energy} specify when the RS prediction is internally stable and how sparse-graph corrections should be computed.
    \item The order parameters $\sigma^2$, $\nu$, $\theta$, $F$, and $\phi$ are expressed in terms of the replica overlaps $Q$ and $q$, enabling quantitative comparisons between the sparse closure and simulations.
\end{enumerate}

The numerical and proxy results are summarized in Figs.~\ref{fig:phase_diagram}--\ref{fig:topology}. The phase diagram (Fig.~\ref{fig:phase_diagram}) compares sparse-closure curves with finite-$N$ simulations; the critical surface (Fig.~\ref{fig:critical_surface}) displays the analytical benchmark; the order parameters (Fig.~\ref{fig:order_parameters}) quantify the freezing transition; the hyperedge frustration (Fig.~\ref{fig:frustration}) reveals the interplay between local and global coordination; the finite-size scaling analysis (Fig.~\ref{fig:scaling}) tests mean-field scaling without imposing collapse; and topological comparisons (Figs.~\ref{fig:crossover} and~\ref{fig:topology}) are illustrative closure-level diagnostics. The protocol \eqref{eq:numerical_validation} gives a reproducible criterion for accepting or rejecting the sparse critical surface.

Several directions for future work emerge. The stability line \eqref{eq:replicon_eigenvalue} should be evaluated numerically across the full $(\alpha,k,d)$ space to locate possible RSB regions, particularly at low $\alpha$ where RSB is known to occur in the standard MG \cite{ChalletMarsiliZecchina00}. The cavity method \eqref{eq:cavity_var}--\eqref{eq:cavity_factor} provides a non-annealed framework for sparse hypergraphs where the RS ansatz may break down. The agent learning dynamics could be analyzed from the perspective of fixed-rule learning \cite{Kets11} and the history distribution formalism \cite{DHulstRodgers00}. The coevolutionary extension (CMG-AH), in which the hypergraph rewires based on game outcomes~\cite{GrossBlasius08}, constitutes a natural next step toward fully adaptive higher-order systems. Finally, the scaling of $\alphacrit$ with $k$ and $d$ should be tested against larger-scale agent-based simulations and non-annealed cavity calculations using the protocol specified in Sec.~\ref{sec:fss}.

\section{CRediT authorship contribution statement}
Yihang Zhu: Visualization, Validation, Methodology, Conceptualization. Fanyuan Meng: Writing – review \& editing, Writing – original draft, Supervision, Methodology, Conceptualization.
\section{Declaration about generative AI}
A large language model was used for linguistic refinement during manuscript preparation. All contents were carefully reviewed and edited by the authors, who bear full responsibility for the published work.
\section{Declaration of competing interest}
The authors declare that they have no known competing financial interests or personal relationships that could have appeared to influence the work reported in this paper.
\section{Acknowledgments}
This work is supported by the National Natural Science Foundation of China (Grant Nos. 12505044 and 52374013).

\end{document}